\documentclass[10pt]{article}
\usepackage[left=1in,right=1in]{geometry}
\usepackage[T1]{fontenc}
\pdfoutput=1
\usepackage{graphicx}
\usepackage{mathptmx}
\usepackage{mathrsfs}
\usepackage{amssymb}
\usepackage{amsmath}
\usepackage{amsmath,amsfonts}
\usepackage{algorithm}
\usepackage{algorithmic}
\usepackage[numbers,square]{natbib}
\usepackage{amscd}
\usepackage[mathscr]{eucal}
\usepackage{tikz}
\usepackage[doublespacing]{setspace}
\usepackage{fancyhdr}
\usepackage{lineno}

\usepackage{enumitem}
\usepackage{stackrel}
\usepackage{url}
\usepackage{etoolbox}

\makeatletter
\patchcmd{\NAT@test}{\else \NAT@nm}{\else \NAT@nmfmt{\NAT@nm}}{}{}

\DeclareRobustCommand\citepos
  {\begingroup
   \let\NAT@nmfmt\NAT@posfmt
   \NAT@swafalse\let\NAT@ctype\z@\NAT@partrue
   \@ifstar{\NAT@fulltrue\NAT@citetp}{\NAT@fullfalse\NAT@citetp}}

\let\NAT@orig@nmfmt\NAT@nmfmt
\def\NAT@posfmt#1{\NAT@orig@nmfmt{#1's}}

\makeatother

\newcommand{\pkg}[1]{\textbf{#1}}
\newcommand{\proglang}[1]{\textsf{#1}}
\newcommand{\code}[1]{\texttt{#1}}

\numberwithin{equation}{section}

\numberwithin{theorem}{section}

\newcommand{\be}{\begin{equation}} \newcommand{\ee}{\end{equation}}
\newcommand{\bd}{\begin{displaymath}} \newcommand{\ed}{\end{displaymath}}
\newcommand{\ben}{\begin{enumerate}} \newcommand{\een}{\end{enumerate}}
\newcommand{\bi}{\begin{itemize}} \newcommand{\ei}{\end{itemize}}
\newcommand{\p}{\parallel}

\newcommand{\ud}{\mathrm{d}}

\newcommand{\cov}[2]{\operatorname{Cov}\left[ #1,#2 \right]}

\newcommand{\Cov}[1]{\operatorname{Var}\left[ #1 \right]}

\newcommand{\Expectation}[1]{\operatorname{E}\left[ #1 \right]}
\newcommand{\Var}[1]{\operatorname{Var}\left[ #1 \right]}
\newcommand{\variance}[1]{\operatorname{Var}\left[ #1 \right]}

\newcounter{rEPP}

\newcounter{rEexpPsi}

\newcounter{rEtauP}

\begin{document}


\title{``Old Techniques for New Times'': the \pkg{RMaCzek} package for producing Czekanowski's Diagrams}

\date{}


\author{{\sc Krzysztof Bartoszek}\thanks{{krzysztof.bartoszek@liu.se, krzbar@protonmail.ch, 
Department of Computer and Information Science, Link\"oping University, 581 83 Link\"oping, Sweden
}}
~and {\sc Albin V{\"a}sterlund}
\thanks{{
abbe--93@hotmail.com, 
Department of Computer and Information Science, Link\"oping University, 581 83 Link\"oping, Sweden
}}}

\maketitle

\begin{abstract}
Inspired by the \pkg{MaCzek} Visual Basic program we provide an \proglang{R} package, \pkg{RMaCzek},
that produces Czekanowski's diagram. Our package permits any seriation and distance method the user provides. 
In this paper we focus on the \code{"OLO"} and \code{"QAP\_2SUM"} methods from the \pkg{seriation} package. 
We illustrate the possibilities of our package with three anthropological studies, one socio--economical one and a 
phylogenetically motivated simulation study.
\end{abstract}


\begin{section}{Introduction} \label{secIntro}

Clustering methods are today a standard way of summarizing even complicated data sets.
Multiple algorithms and approaches are possible \citep[for an introductory overview see e.g.][]{JKog2007},
starting with the classical k--means algorithm \citep{HSte1956}.
Alternatively to visualizing the originally  measured values one may present observations
using the distances between them, e.g. through multidimensional scaling plots 
\citep[e.g. Ch. 4,][]{BEveTHot2011}.
All of these are possible due to the advancements in 
digital electronic computers---permitting rapid calculations and high--resolution colour
presentation of data. This was not the case always. At the beginning of the previous century
high--dimensional data was also collected but its presentation was restricted to usually 
black--and--white, manually created figures on paper. With these constraints, the Polish
anthropologist \citet{JCze1909} proposed the now--called Czekanowski's Diagram. 
To construct such a diagram one only needs to be able to calculate the distance
between observations. Then, the next step is to find an ordering of the data points, so
that those with small distances are close to each other, i.e. a seriation 
problem \citep{MHahKHorCBuc2008}.
Finally, one plots a matrix (rows and columns
correspond to observations), where each cell contains a symbol representing the distance
between the appropriate pair of objects. As the number of symbols is limited, the 
distances have to be grouped, e.g. the interval $($minimum distance, maximum distance$)$
is split into a number of consecutive subintervals and each interval gets assigned 
a unique symbol (e.g. dots with varying sizes, see e.g. Fig. \ref{figphylQAP2SUMOLOGA}
or \ref{figJCze1909}).
Even though time--consuming, it was possible to manually present clusterings
of high--dimensional observations, provided that the number of data points is not 
overly large. For example, \citet{JCze1909} had distances between $13$ 
skulls belonging to various archaic humans. Each skull was characterized by
$27$ variables, but there were missing measurements
\citep[original data from][]{KSto1908}. 
The largest manually ordered diagram, that we are aware of, consists of 
$108$ observations.
\citet{ASolPJas1999} report that such diagrams can be found in the archives of Warsaw University.
These large diagrams, carefully done with a pen, are in the documents left
by Boles{\l }aw Rosi{\'n}ski, a student of Jan Czekanowski.  
Andrzej Wierci{\'n}ski remembered (in discussions with Arkadiusz So{\l }tysiak)
that as a young student in $1940$s and $1950$s he spent a lot of time on arranging and drawing these diagrams
(personal communication with Arkadiusz So\l tysiak, $19$ III $2019$).

However, the advent of computers allowed these operations to be done automatically.
In the $1980$s two programs are reported 
to have appeared
(\url{http://www.antropologia.uw.edu.pl/MaCzek/maczek.html}), \pkg{Ediaczek}, for botanists
\citep[has been reported in use by][]{ACie2009}
and a program of F. Szczotka
(\citet{FSzc1972} and also \citet{PBer2003}
reported personal communication).
Then, \citet{ASolPJas1999} developed, an originally \proglang{Turbo Pascal}, and later
\proglang{Visual Basic} program \pkg{MaCzek} 
that creates Czekanowski's diagram and allows the user, amongst others, to choose the distance function, seriation
method (including manual rearrangement) and symbols for distance intervals
(\url{http://www.antropologia.uw.edu.pl/MaCzek/maczek.html}). 
\citet{ILii2010} provides a review and historical perspective on these methods
and \citet{PJasASol2004} discuss the development and various applications,
with an interesting focus on the pre--computer era, of Czekanowski's diagram
(unfortunately for most readers, in Polish).
Even though \pkg{MaCzek} seems to be the most
popular program to create these diagrams it 
seems to be confined to 
(predominantly Polish) anthropological \citep[e.g.][]{ASol2000},
botanical \citep[e.g.][]{ACie2009}, environmental \citep[e.g.][]{ABorBSkwGOls2012,SDolZMigAMic2013}
and socio--economical \citep[e.g.][]{MHomMMos2016,AKra2009,KWar2015} communities.
This can be attributed to \pkg{MaCzek}'s webpage 
and manual
being provided only in Polish. Furthermore, \pkg{MaCzek} only runs on Windows operating systems
and can handle only up to $250$ observations with up to $100$ variables. 
With contemporary big data and multitude of software platforms such restrictions seem
prohibitive. Following encouragement from Miros{\l }aw Krzy{\'s}ko and
Arkadiusz So{\l }tysiak \citet{AVas2019} provided an \proglang{R} implementation of
creating Czekanowski's diagram through the \pkg{RMaCzek} package 
(available on CRAN \url{https://cran.r-project.org/web/packages/RMaCzek/}). 
An important component of this \proglang{R} implementation is that any user
provided distance and seriation method can be used.
By default \pkg{RMaCzek} offers the \code{stats::dist()} method, 
all (applicable) methods from the \pkg{seriation} package and
a custom implemented genetic algorithm. However, 
the user may provide their own distance function and 
using the interface offered by the
\pkg{seriation} package a user may use their own custom arrangement function,
see Section \ref{sbsecSeriation}.
Furthermore, the number of observations and their dimension are only
constrained by the user's memory, CPU and graphical display capabilities.

We hope that \pkg{RMaCzek}'s \proglang{R} implementation will popularize 
one of the first clustering/cluster visualization/taxonomic methods
that has at its core simplicity and compactness of presentation. 
In today's world it might seem at first that there is a nearly 
infinite capacity for displaying information and that one should 
rather aim at presenting more and more complexity. 
However, due to its inherent simplicity, one of the main advantages, 
over today's popular heatmaps, dendrograms
and other highly sophisticated approaches, of the diagram seems to
be that at one glance not only the groupings can be visualized
but also the relationships between and inside the groupings. 
An interesting example of this is given by \citet{PJasASol2004}. They
consider $1993$ economical data from ``old''--EU countries. Two clusters
can be seen from a dendrogram 
(low GDP and high percentage of rural population: Greece, Portugal, Ireland and Spain; 
high GDP and low percentage of rural population:
Belgium, Holland, Great Britain, France, Germany, Italy, Denmark and Luxembourg).
Czekanowski's diagram produces the same clusters---but furthermore shows
that Spain is on the boundary of these two clusters---completely
missed by the dendrogram.
Furthermore, one should not overlook the existence of situations where a low--weight
graphical representation is required. Examples of these could be mobile webpages,
low bandwidth connections, low resolution graphical displays,
grayscale journal requirements or sometimes even an overload of colours/detail
can hamper human perception. In all such cases a method designed for manual, analog
cluster creation and presentation transferred to a digital, operating system independent
(as long as \proglang{R} permits) platform could be the answer. Paraphrasing the name
of the conference where \citet{ASolPJas1999} presented their \pkg{MaCzek} program---
``Old Techniques for New Times''.

The paper is arranged as follows, we first describe what is 
Czekanowski's diagram (Section \ref{secCzkDiag}),
then in Section \ref{secObj} discuss what is being optimized, with an example simulation 
study. Afterwords, in Section \ref{secRMaCzek} we provide details about 
the \pkg{RMaCzek} \proglang{R} package. We end with a number of example 
analyses in Section \ref{secAnalyses}. 
They are reruns of our package on data from which
Czekanowski's diagrams were previously constructed, for comparison of our software with other 
implementations. We end with a short Discussion.

\end{section}

\begin{section}{Czekanowski's diagram}\label{secCzkDiag}
It would seem \citep{ILii2010} that \citepos{JCze1909} approach ``was the 
first published work on one--mode data analysis that was based on the permutation
of the rows and columns, complemented with color (pattern) coding for better 
visual perception.'' As already written in the Introduction the goal of \citet{JCze1909}
was to create a matrix that best visualizes 
the (dis)similarities between a collection of observations. 
As a picture is worth a thousand words, the reader is referred to
Figs. \ref{figphylQAP2SUMOLOGA} or \ref{figJCze1909} to see what Czekanowski's
diagram is. 
The entries of the plotted on the diagram matrix
should correspond to the (dis)similarity between the appropriate row and column
observation. However, instead of writing/colour--coding the exact distance/similarity
function value one replaces these ``with simple graphic characters, e.g. white squares
filled with black rectangles or black dots, of differentiated diameter'' \citep{ASolPJas1999}. The most similar
to each other objects should have the biggest/most noticeable objects assigned while
the least can have nothing---a purely white square. As \citet{ASolPJas1999} point out
that changing range of (dis)similarity values assigned to a character, changing the number or type of 
characters one ``can easily emphasize, or blur, some tendencies.'' The exact choice
of how to divide the distance values into common symbols can be a subjective 
procedure \citep{PBer2003} up to the user's discretion. However, the \pkg{RMaCzek} package
offers the possibility of a more objective approach. The user may declare what percentage of the 
distances falls into each group (default all groups have to be equal).
We feel that this is closer to \citepos{JCze1909} original proposal, which is also implemented
and described towards the end of this Section.

An important aspect of the diagram
is its potential for simplicity. The discrete number of symbols allows for easy visualization 
of the cluster structure at 
the user preferred level of detail. This level of abstraction is controlled by the number of used
symbols---the more, the more detailed picture but fewer generalizations are directly presented.

It is important to point out that to create such a diagram one needs only to be able to calculate
a distance/similarity value between a pair of observations. \citet{ASolPJas1999}
wrote that all the variables describing an observation must be of the same type
(e.g. numeric, ordinal). This was certainly true, with the original distance proposed
by \citet{JCze1909}, the average difference (Manhattan distance divided by the number of variables),
or the Euclidean distance. In fact \citepos{JCze1909} approach was most criticized for the 
fact that his originally proposed metric did not account for scale effects nor correlations between
the variables \citep{PBer2003}. 
However, this should not be considered to be a fault of the proposed visualization method.
Rather it is a ``degree--of--freedom'' situation where the user should 
appropriately pre--process the data and choose an appropriate
distance function (\pkg{RMaCzek} package allows the user to specify their own) 
that can e.g. handle different variable types. 
For example to take care of scale (and location) effects the data can be 
standardized prior to any further steps 
(this is a user controlled option in \pkg{RMaCzek}), or correlations can be taken into account by 
using e.g. the Mahalanobis distance.
Furthermore, missing values for some
observations can be tolerated as long as the distance function can handle this.
However, as \citet{ASolPJas1999} warn, one must be careful that an observation
with multiple missing values can be close to multiple very different objects.
Hence, they suggest, based on their experience, observations with more
than $50\%$ missing variables should be removed from further analyses.

Given, the distances between the observations comes the question how 
to arrange them in the rows (and in the same way in the columns). 
We would desire similar observations to be close to each other in the ordering, while
dissimilar far away. \citet{ILii2010} writes that 
``Czekanowski did not have any formal procedure for rearrangement of the
elements in the matrix; therefore, probably, visual inspection and intuition was used because the size of the dataset
was also considerably small.''
\citet{JCze1909} does not mention how he ordered the skulls but
when one carefully inspects \citepos{JCze1909} Tabs. II and III,
and his contemporary literature then a plausible justification can be found. 
\citet{KSto1907} reports, after \citet{GSch1906}, the following evolutionary
timeline (of archaic humans) based on anatomical relationships (cf. Fig. \ref{figJCze1909}):
Neandertal, Spy, Krapina 
(\citet{GSch1906} and \citet{KSto1907} classified the last three as H. primigenius),
Gibraltar, Br\"ux, Galley--Hill, Br\"unn
(\citet{GSch1906} and \citet{KSto1907} classified the last three as H. sapiens),
with Pithecanthropus earlier than all of them, but uncertain if an ancestor.
This ordering, corresponds with minor rearrangements, to 
\citepos{JCze1909} one. Furthermore, one needs to add the Kannstatt
and Egisheim skulls. The final, Nowosio{\l }ka skull, is a focal observation
whose placement was \citepos{JCze1909} goal.
The skull was then hypothesized, by \citet{KSto1908} to be related 
(based on an interpretation of morphological comparisons) with the Krapina--Spy--Neandertal
group. However, \citepos{JCze1909} taxonomic procedure, suggests that it is 
rather closer related to the ``H. sapiens cluster''. In Section \ref{secAnalyses} 
we reanalyze the data using \pkg{RMaCzek} and find the same.

However, usually today's researcher will not have such in--depth knowledge, from years of scientific debate,
on the observations and contemporary data collections are significantly larger, making
manual arrangements next to impossible.
Furthermore, computers, as of yet, are not able to apply
human intuition in such a context. Therefore, algorithmic approaches are necessary.
These fall under the generic name of seriation and matrix reorder methods.
Seriation is a way or reordering the sequence of observations (along a one--dimensional line)
so that patterns and regularities in them are best revealed \citep{ILii2010}. 
These patterns can be found on the between--individual level, between groups of objects and 
on the global level \citep{ILii2010}. Furthermore, because the result is an ordered sequence
of observations along a line one can observe the chain of associations between the 
objects \citep{ILii2010}. 

There are multiple ways of seriation, \citet{ILii2010} provides a historical overview
of them and e.g. the \proglang{R} package \pkg{seriation} 
\citep{MHahKHorCBuc2008} implements a number of them. The \pkg{MaCzek} program
offers three seriation approaches. From its description, a simple one where 
the two most similar objects are joined at each step and two possible 
genetic algorithms. The implementation of a genetic algorithm requires
the implementation of an objective function that scores how good 
a permutation is with respect to the distance matrix. Such an objective
function should encourage similar objects to be close to each other,
while distant objects to be far away from each---in effect producing
a diagram that looks like a series of hills with peaks along the matrix's diagonal.
The main clusters are arranged along the diagonal, and inside each cluster we want
similar diagonal arrangements. \citet{ASolPJas1999} proposed $U_{m}$, Eq. \eqref{eqUm},
as such a function. In Section \ref{secObj} we investigate it and consider an alternative
approach.

It is important to underline that the seriation of the observations does not depend on
the discretization of the distances. The optimal ordering is found based on the
original distances, and the assignment of the symbols representing the distance
classes takes place only afterwords, at the visualization stage.

Finally, all of Czekanowski's diagrams that we came across in the literature are 
symmetric---each cell is a direct encoding (by the chosen symbols) of the 
corresponding entry in the distance matrix. However, this is not what
\citet{JCze1909} originally suggested. \citet{JCze1909} treated each
column of the distance matrix separately. Each column contains
distances to a given object (let us call it the ``column object''). 
For each column he assigned
one symbol to the three most similar observations to the ``column object'',
then the fourth most similar obtained the next (remember there should be some natural ordering 
to the symbols in order to perceive the similarity)  symbol, the fifth most similar 
the next symbol and the final symbol was assigned to the sixth most similar.
All the other cells in the column obtained a blank symbol. 
Then, similarity between the observations was assessed on the similarity
of their column patterns. If two objects are similar to the same objects, then 
they are understood to share a lot between themselves, despite perhaps not being
directly similar.

In fact \citet{PBer2003} writes that in his personal opinion assigning separate 
symbols column by column to the three most similar, than separately the fourth, fifth,
sixth and leaving the others blank is more objective than subjectively
grouping distances into the same symbol.
In this spirit the \pkg{RMaCzek} package allows the user to choose how to group
the ranks of the similarities in each column (same grouping for each column).
By default \citepos{JCze1909} original grouping $\{\{1,2,3\},\{4\},\{5\},\{6\},\{\mathrm{rest}\}\}$
(where $i$ means the $i$--th most similar) is used. However, one must also remember
that \citet{JCze1909} proposed his grouping based on his visualization possibilities.
Therefore, today's graphical possibilities allow for possibly
more appropriate groupings for a given dataset.

Up to now we have discussed clusters in the data without actually saying
how to identify them. The matrix permutation algorithms have an advantage
when compared with cluster algorithms in that 
``no information of any kind is lost, and that the number of clusters
does not have to be presumed; it is easily and naturally visible''
\citep[p. 212][]{HSpa1980}. While visually inspecting Czekanowski's diagram
clusters can be clearly visible, an additional step is required
to specify their boundaries. This could of course be a manual cut
but otherwise some sort of computational procedure is required. 
Implementing and proposing such methods is beyond this work
but an immediate approach could be to use a dendrogram construction
procedure, that does not re--arrange the objects. Then, 
cluster membership would be taken according to appropriate number of top splits,
so that the desired/observed number of clusters is obtained.
In Section \ref{secObj} we see how well \pkg{RMaCzek}'s different methods
are able to identify the main clades for phylogenetically correlated data.
A similar approach has been formally proposed by \citet{FSzc1972}
based on the so--called Wroc{\l }aw taxonomic method \citep{KFloJLukJPerHSteSZub1951}.
Furthermore, \citet{THavJBezJKelMPop2009} also consider clustering of ordered distance matrices.
However, their visualization is symmetric, restricted to grayscale encoding of
the distance matrix and no implementation seems to be provided.
\end{section}

\begin{section}{The objective function}\label{secObj}
As already mentioned in Section \ref{secCzkDiag} in order to automatically re--order 
the observations one needs some sort of seriation algorithm.
In their original implementation, \citet{ASolPJas1999} scored
a permutation, $\pi(\cdot)$, of the objects according to the formula

\be\label{eqUm}
U_{m} = \frac{2}{n^{2}}\sum\limits_{j=1}^{n}\sum\limits_{i=j+1}^{n}\frac{(i-j)^{2}}{W_{\pi(i),\pi(j)}+1},
\ee
where $W_{ij}=d(a_{i},a_{j}$) is the distance between objects at positions $i$ and $j$.
The lower the value of $U_{m}$, the better the clustering. 
The number of possible permutations is $n!$ hence considering all possible and then choosing 
the one with the lowest value of $U_{m}$ is only possible for very small datasets.
Therefore, a genetic algorithm optimization method, based on \code{GA::ga()} \citep{LScr2013,LScr2017},
that searches for the permutation with minimum $U_{m}$,
is shipped with \pkg{RMaCzek}.
Furthermore, seriation methods found in the \pkg{seriation} package
are also available with alternative objective functions. 

\citet{AVas2019} found empirically that 
minimizing $U_{m}$ might not be the best strategy. In the course of his experiments
it was observed that the \code{"QAP\_2SUM"} \citep{SBarAPotHSim1993}
consistently best minimizes $U_{m}$ for a given dataset. 
This is not surprising as, based on the manual pages of
\code{seriation::criterion()} and \code{seriation::seriate()},
\code{"QAP\_2SUM"} minimizes, over permutations, $\pi(\cdot)$, of $(1,\ldots,n)$ the
so--called $2$--sum criterion

\be
L_{2sum} = \sum\limits_{j=1}^{n}\sum\limits_{i=1}^{n}\frac{(i-j)^{2}}{W_{\pi(i),\pi(j)}+1}.
\ee
Hence, as distances are symmetric, it is obvious that $L_{2sum} = n^{2}U_{m}$ and
minimization of these two objectives is equivalent to each other.
However, it was also observed \citep[subjective evaluation of resulting Czekanowski's diagrams by][]{AVas2019}
that \code{"QAP\_2SUM"} minimization (and so $U_{m}$) tends to work by forcing objects that have a large distance
between each other to be far away. This might not be the best as close--by objects might not be clustered together.
On the other hand the \code{"OLO"} \citep[Optimal leaf ordering, ][]{ZBaretal2001} method consistently 
did not perform best at minimizing $U_{m}$,
not surprisingly as its optimization goal is something else,
but produced visually more appealing diagrams. Citing the help page of
\code{seriation::seriate()}, \code{"OLO"}
``produces an optimal leaf ordering with respect to the
minimizing the sum of the distances along the (Hamiltonian)
path connecting the leaves in the given order.''
More formally \citep{DEarCHur2015}, the permutation $\pi(\cdot)$ of the $(1,\ldots,n)$ objects is found
such that 

\be
L_{\mathrm{path~length}} = \sum\limits_{i=1}^{n-1} W_{\pi(i),\pi(i+1)}
\ee
is minimized. Notice that this corresponds to finding the minimal length
Hamiltonian (each node visited exactly once) path inside the complete graph
where each vertex is an observation and branch between two vertices
has length equalling the distance between the two corresponding objects. 
Finding such a permutation is NP--hard, as it the same as solving
the travelling salesman problem (without returning to the origin). 

From \citepos{AVas2019} numerical experiments it turns out that, from our perspective, the 
path length minimization method's main focus is
on clustering close--by objects. And it seems that in turn a better Czekanowski's diagram is produced.
We compare these three methods in 
Tabs. \ref{tabphylQAP2SUMOLOGA_clusters}, \ref{tabphylQAP2SUMOLOGA_Um}, \ref{tabphylQAP2SUMOLOGA_OLO} 
and provide an example visualization
of a dataset illustrating that minimizing $U_{m}$, 
might not provide the best ordering 
and minimization of path length works better.

We simulate phylogenetically correlated data on a tree
that has three distinct clades (clusters). The phylogenetic tree was simulated using
the function
\code{mvSLOUCH::simulate\_clustered\_phylogeny()},
from the package \pkg{mvSLOUCH} \citep{KBarJPiePMosSAndTHan2012}. It simulates 
a given number of pure birth trees \citep[using \pkg{TreeSim}][]{TreeSim1,TreeSim2}
and then joins them by a pure birth tree. However, the ``tip branches'', i.e. those leading to
the cluster, of the joining phylogeny are elongated---hence causing the clades to be distant
from each other, see Fig. \ref{figphylQAP2SUMOLOGA}. Then, on top of the phylogeny a 
$10$--dimensional Ornstein--Uhlenbeck (OU) process is simulated, using \pkg{mvSLOUCH}, 
with randomly generated coefficients, see the
code in Supplementary Material. We evaluate, over a $100$ repeats, how all the three methods 
minimize $U_{m}$ (Tab. \ref{tabphylQAP2SUMOLOGA_Um}), the path length criterion (Tab. \ref{tabphylQAP2SUMOLOGA_OLO}), 
how correct the clusters are
(Tab. \ref{tabphylQAP2SUMOLOGA_clusters}) and provide a visualization of the study and example Czekanowski's diagrams
for the setup (Fig. \ref{figphylQAP2SUMOLOGA}). Our phylogeny has three distinct clades of size $30$ tips each. 
Each clade has an ``optimal value'' of the OU process.
If these optima are clade specific, then this should make the clades more distinctive.
The trait values in the clades would then exhibit similarity not only through greater evolutionary relatedness
(stronger correlations) but also through common to the clade mean value. 
The actual values of the optima are randomly drawn from a normal distribution.
We consider three scenarios. The three clades have optima drawn 
from Gaussians with very different means (labelled distinct optima), different but similar means
(labelled similar optima) or with equal means (labelled ``equal'' optima).
This is seen in the different ways \code{mPsi} is set in 
function calls below.
The matrix of trait values has rows ordered according to the phylogeny---hence the \code{czek\_matrix()} starts with the 
correct ordering of the observations. We also study how it behaves if the traits' measurement matrix's
rows are randomly shuffled. We scale the data, i.e. \code{scale\_data=TRUE}, so the actual values of the optima
do not matter, only how different they are from each other. 

To assess correctness of the clusters we report in 
Tab. \ref{tabphylQAP2SUMOLOGA_clusters} how many tips from each cluster are in the first, second and 
third thirty observations. As there are multiple repeats, we make a final rearrangement of the ordering
that minimizes the mismatch for the three groups (function \code{f\_cluster\_assess()}, see code in Supplementary Material).
The data for Tab. \ref{tabphylQAP2SUMOLOGA_clusters} and Fig. \ref{figphylQAP2SUMOLOGA}
were produced by the calling the \code{f\_doRMaCzekAnalysis()} in the \proglang{R} code in the Supplementary Material.

The function \code{mvSLOUCH::simulate\_clustered\_phylogeny()} 
enhances the usual \code{phylo} class object with two extra fields \code{edges\_clusters}
(to which cluster does each edge belong or to the joining part of the tree)
and \code{tips\_clusters} (to which cluster does each tip belong to). This enhanced object then
belongs to the custom \code{"clustered\_phylo"} class. 
Then, such a phylogeny can be plotted directly using \pkg{mvSLOUCH}'s capabilities,
\\ \noindent
\code{> phyltree<-output[[1]]\$phyltree}\\ \noindent
\code{> plot(output[[1]]\$phyltree,clust\_cols=c("blue","green","red"),}\\  \noindent
\code{		clust\_edge.width=3,joiningphylo\_edge.width=3,}\\  \noindent
\code{		show.tip.label=FALSE)}\\  \noindent
where the \code{plot()} function for a \code{"clustered\_phylo"} class 
is based on \code{ape::plot.phylo()} \citep[from the \pkg{ape} package,][]{EParKSch2018}.

We can see in Tab. \ref{tabphylQAP2SUMOLOGA_clusters} that the previously mentioned observations
seemed correct. First in the easiest case, when the optima
of the three clusters are distinct, all three methods produced a completely/nearly correct 
clustering. When the optima became more similar, then the three methods behaved similarly, clustering
on average $80\%$ objects correctly but it would seem that the optimal leaf ordering
has a slight edge. However, optimal leaf ordering completely outperforms the other two methods
in the most difficult setup---when the optimal values for the clusters are drawn from the same distribution
and also the ordering of the input data was shuffled. 
The results in Tabs.  \ref{tabphylQAP2SUMOLOGA_Um} and \ref{tabphylQAP2SUMOLOGA_OLO}
confirm that all three methods optimize what they are meant to optimize.
The genetic algorithm and \code{"QAP\_2SUM"} minimizes $U_{m}$ better 
than \code{"OLO"}, while the latter minimizes better than the two former, the path length
criterion. This pattern is visible in all the setups, both those ``easy'' and ``difficult''.
However, \code{"OLO"} is only slightly worse in minimizing $U_{m}$ but it completely
outperforms the other methods in minimizing the Hamiltonian path length. 
This is again evident in both the ``easy'' and ``difficult'' setups. 
Table \ref{tabphylQAP2SUMOLOGA_timing} illustrates another big advantage
of the \code{"OLO"} method--it is faster than \code{"QAP\_2SUM"}.
The custom \pkg{RMaCzek} genetic algorithm has an extremely long running time and given
that it does not perform any better than \code{"OLO"} and  \code{"QAP\_2SUM"}, it is kept only as a backup so that the 
package has its own internal seriation method.
In Section \ref{secAnalyses} we run the \code{"OLO"} method on datasets that were arranged 
by \pkg{MaCzek} and it would seem that while both capture 
the same clusters, path length minimization seems to be able to better arrange inside 
and between the clusters. 
This indicates that path length minimization seems to better captures the transitions
between and inside clusters.

\begin{table}[p]
\caption{
Fraction out of $100$ repeats of correctly clustered observations by the \texttt{czek\_matrix()} function
for different simulation setups. In brackets the standard deviation of the fraction.
Each cluster (clade on the phylogeny) had $30$ observations (tips). The ``Equal'' optima
does not mean that the optima were actually equal---only that they were drawn from a normal 
distribution with the same expected value. 
\label{tabphylQAP2SUMOLOGA_clusters} 
}
\begin{center}
\begin{tabular}{cccc}
\hline
Simulation type & \code{order="ga"} &  \code{order="QAP\_2SUM"} & \code{order="OLO"} \\ \hline
Correct input order &  &  &  \\ \hline
Distinct optima & $1,1,1$ & $1,1,1$ & $1,1,1$ \\
& $(0),(0),(0)$ & $(0),(0.003),(0.003)$ & $(0),(0),(0)$ \\
Similar optima & $0.94,0.882,0.941 $ & $0.894,0.826 ,0.898$ & $0.947, 0.887, 0.931 $ \\
& $(0.092) ,(0.115),(0.085)$ & $(0.119),(0.135) ,(0.115)$ & $(0.09) ,(0.132) ,(0.119)$ \\
``Equal'' optima & $0.776 ,0.657,0.648$ & $0.678 ,0.634,0.58$ & $0.745 ,0.707 ,0.659 $ \\
& $(0.18), (0.17),(0.152)$ & $(0.16),(0.172),(0.144)$ & $(0.175),(0.174),(0.181)$ \\
\hline \\ \hline
Shuffled input order&  &  &  \\ \hline
Distinct optima & $1,1 ,1 $ & $1,1 ,1 $ & $1,1,1 $ \\
& $(0), (0.003), (0.003)$ & $(0.011), (0.011), (0)$ & $(0),(0),(0) $ \\
Similar optima & $0.832, 0.739, 0.873$ & $0.853, 0.765, 0.879 $ & $0.891, 0.833, 0.901 $ \\
& $(0.136), (0.14), (0.12)$ & $(0.136) ,(0.137) ,(0.124)$ & $(0.133),(0.161) ,(0.139)$ \\
``Equal'' optima & $0.599, 0.567, 0.603 $ & $0.606,0.572,0.605$ & $0.681,0.659,0.673 $ \\
& $(0.151) ,(0.15),(0.161)$ & $(0.152),(0.144),(0.171)$ & $(0.179),(0.163),(0.187)$ 
\\ \hline
\end{tabular}
\end{center}
\end{table}

\begin{table}[p]
\caption{ 
Average and standard deviation of the final values of $U_{m}$ associated with Czekanowski's
diagram proposed by the different methods. \label{tabphylQAP2SUMOLOGA_Um}
}
\begin{center}
\begin{tabular}{cccc}
\hline
Simulation type & \code{order="ga"} &  \code{order="QAP\_2SUM"} & \code{order="OLO"} \\ \hline
Correct input order & & & \\ \hline
Distinct optima & $218.867(5.552)$ & $219.578(13.966)$ & $218.461(5.735)$ \\
Similar optima & $221.248(4.807)$ & $217.471(4.456)$ & $230.426(7.438)$ \\
``Equal'' optima & $230.268(3.759)$ & $223.779(2.777)$ & $240.116(4.972)$ \\
\hline \\ \hline
Shuffled input order & & & \\ \hline
Distinct optima & $218.956(6.002)$ & $220.324(15.644)$ & $219.412(5.757)$ \\
Similar optima & $223.693(4.798)$ & $218.567(3.587)$ & $233.256(7.486)$  \\
``Equal'' optima & $231.526(3.803)$ & $224.539(2.8)$ & $240.827(4.683)$ 
\\ \hline
\end{tabular}
\end{center}
\end{table}
 
\begin{table}[p]
\caption{
Average and standard deviation of the final values of \texttt{"Path\_length"} criterion
(as returned by \texttt{seriation::criterion()}) associated with Czekanowski's
diagram proposed by the different methods.\label{tabphylQAP2SUMOLOGA_OLO}  
}
\begin{center}
\begin{tabular}{cccc}
\hline
Simulation type & \code{order="ga"} &  \code{order="QAP\_2SUM"} & \code{order="OLO"} \\ \hline
Correct input order & & & \\ \hline
Distinct optima & $57.593(22.427)$ & $46.197(17.772)$ & $33.735(10.353)$ \\
Similar optima & $262.733(20.803)$ & $263.106(23.578)$ & $161.437(9.774)$ \\
``Equal'' optima & $305.577(19.34)$ & $306.774(17.704)$ & $184.665(10.194)$ \\
\hline \\ \hline
Shuffled input order & & & \\ \hline
Distinct optima & $52.086(17.958)$ & $43.672(14.887)$ & $32.666(9.213)$ \\
Similar optima & $278.27(21.672)$ & $272.043(23.911)$ & $167.347(12.091)$  \\
``Equal'' optima & $316.262(17.672)$ & $311.837(17.619)$ & $189.062(12.371)$ 
\\ \hline
\end{tabular}
\end{center}

\end{table}

\begin{table}[p]
\caption{
Average and standard deviation of running times in seconds for the simulation results presented in
Tab. \ref{tabphylQAP2SUMOLOGA_clusters}. The simulations were run using \textbf{RMaCzek 1.2.0} 
in \textsf{R 3.6.1}  on 
a $3.50$GHz Intel\textsuperscript{\textregistered} Xeon\textsuperscript{\textregistered} CPU. 
While absolute times might not be comparable as other tasks were also running on the machine, the 
relative times immediately show that the genetic algorithm's running time is 
prohibitively large in comparison to the two other methods from the \textbf{seriation} package.
\label{tabphylQAP2SUMOLOGA_timing} 
}
\begin{center}
\begin{tabular}{cccc}
\hline
Simulation type & \code{order="ga"} &  \code{order="QAP\_2SUM"} & \code{order="OLO"} \\ \hline
Correct input order & & & \\ \hline
Distinct optima & $506.275(5.529)$ & $0.0613(0.0214)$ & $0.041(0.013)$ \\
Similar optima & $498.705(19.066)$ & $0.057(0.015)$ & $0.039(0.011)$ \\
``Equal'' optima & $440.94(4.096)$ & $0.052(0.01)$ & $0.036(0.006)$ \\
\hline \\ \hline
Shuffled input order & & & \\ \hline
Distinct optima & $498.969(7.916)$ & $0.056(0.016)$ & $0.04(0.013)$ \\
Similar optima & $506.927(4.427)$ & $0.06(0.03)$ & $0.046(0.026)$ \\
``Equal'' optima & $443.235(13.664)$ & $0.055(0.012)$ & $0.037(0.006)$ 
\\ \hline
\end{tabular}
\end{center}
\end{table}

\begin{figure}[p]
\begin{center}
\includegraphics[width=0.45\textwidth]{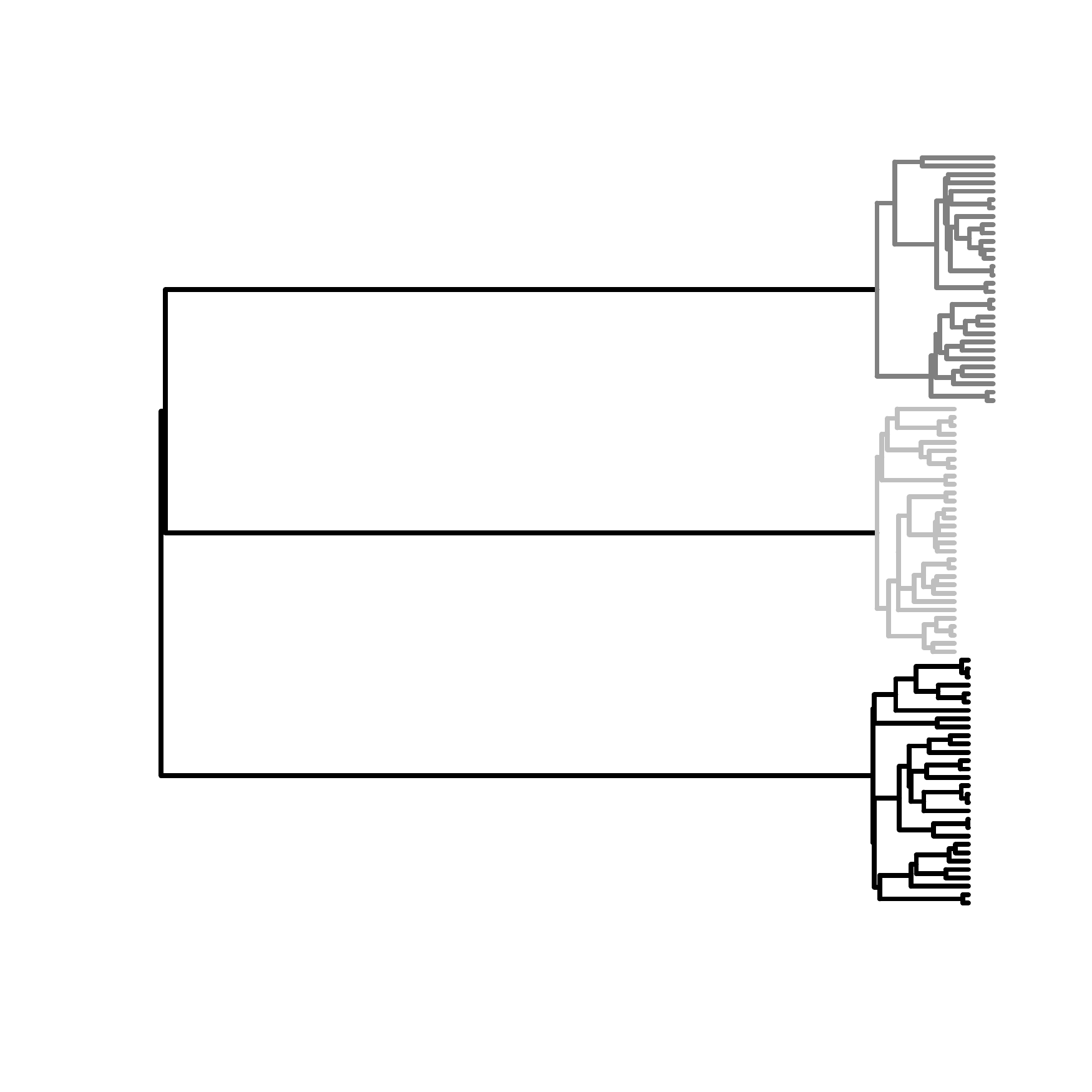}
\includegraphics[width=0.45\textwidth]{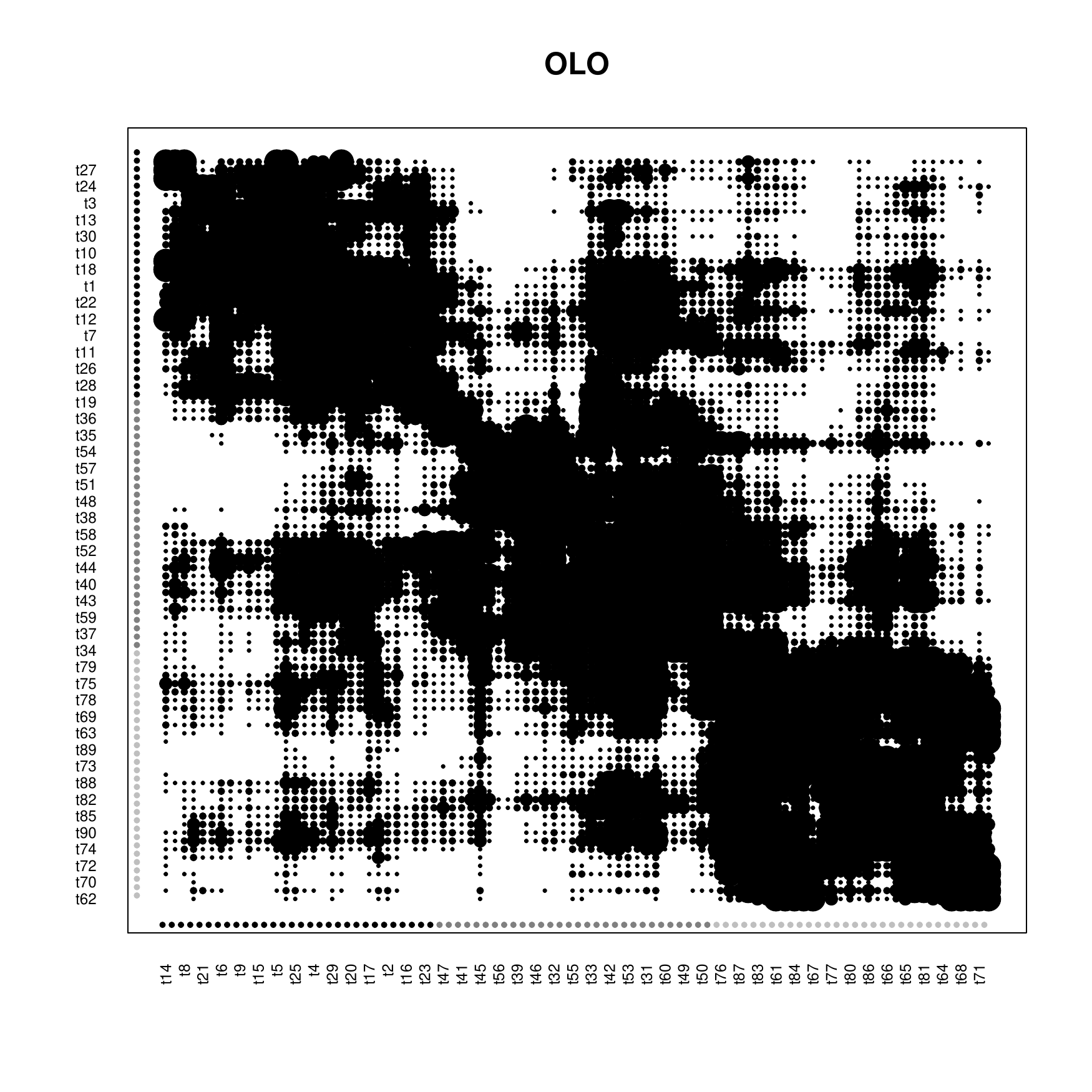} \\
\includegraphics[width=0.45\textwidth]{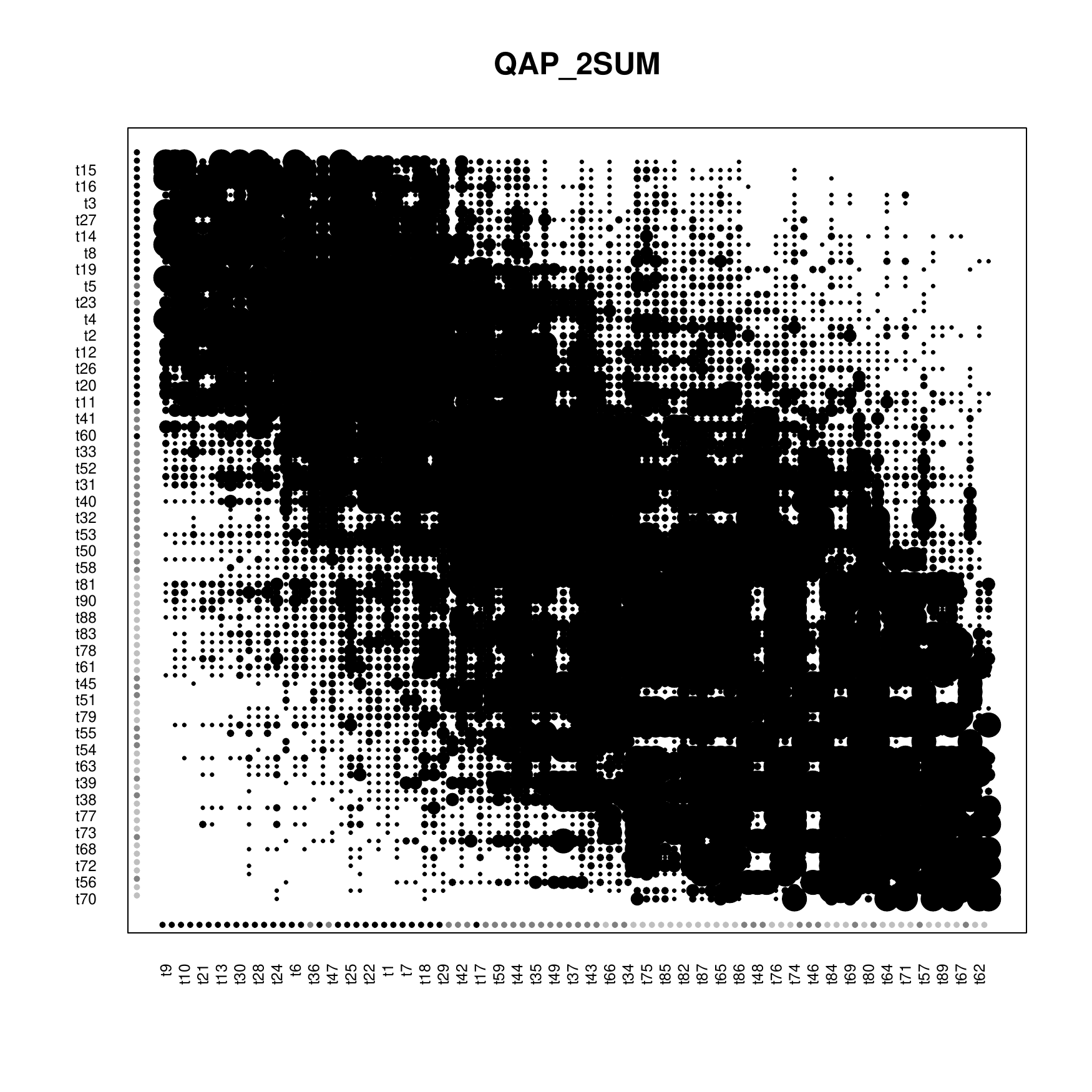}
\includegraphics[width=0.45\textwidth]{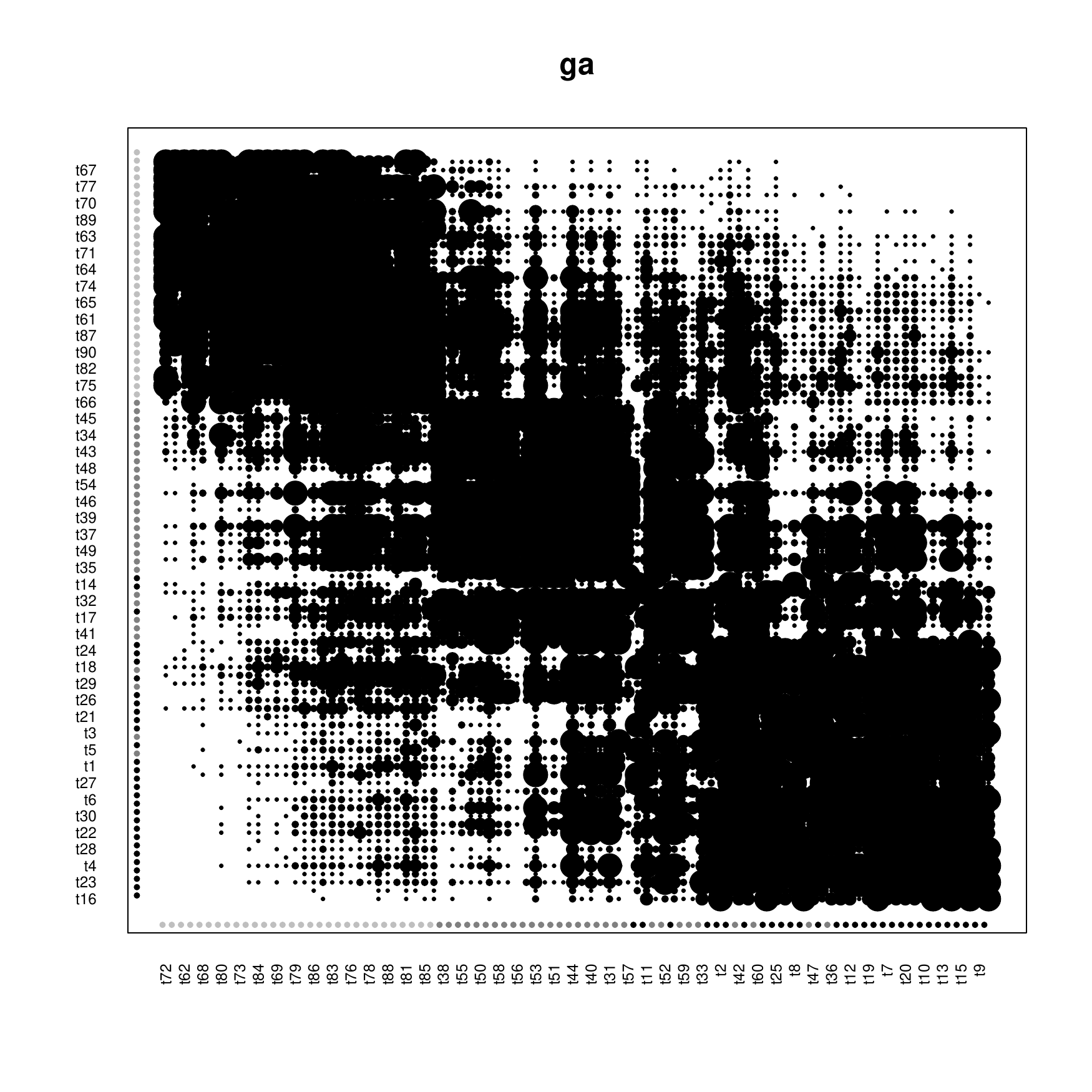} 
\end{center}
\caption{ Example clustered phylogeny and Czekanowski's diagrams
from a $10$--dimensional OU process simulated on the tree. The optima values are the same
as in the similar means setup. The input data is not shuffled, i.e. the values 
are passed in the correct ordering, prior to the seriation. The colours of the dots along the vertical and
horizontal axes of the diagram correspond to the cluster to which the given observation belongs to.
Top right: diagram produced by \texttt{order="OLO"}, bottom left: by \texttt{order="QAP\_2SUM"}
and bottom right: by \texttt{order="ga"}. We can see that \texttt{order="OLO"} recovers the
clusters correctly unlike the two other methods. It has to be pointed out that this figure
is for illustrative purposes and a number of reruns
were required to obtain the above graphs, usually \texttt{"OLO"} would also have some
mistakes, sometimes \texttt{"QAP\_2SUM"} and \texttt{"ga"} would do better
than seen above. However, it was never (visually) observed for \texttt{"OLO"} to do worse than
\texttt{"QAP\_2SUM"} or \texttt{"ga"}.
\label{figphylQAP2SUMOLOGA}
}
\end{figure}
\end{section}

\begin{section}{The RMaCzek package}\label{secRMaCzek}
\begin{subsection}{User interface}\label{sbsecUI}
The \pkg{RMaCzek} package offers the user two main functions
\code{czek\_matrix()} and \code{plot.czek\_matrix()}. The first one 
takes the user's data (in the form of a numeric matrix,
data frame or matrix of distances between observations, i.e. a \code{"dist"} object in the last case)
and returns a matrix of distances between the observations
(of class \code{"czek\_matrix"}). The user is able
to specify, amongst others, the seriation method to be used
(parameter \code{"order"}), 
the number of classes (parameter \code{"n\_classes"})
the intervals (parameter \code{"interval\_breaks"}),
and the distance function to be used (parameter \code{"distfun"}, default \code{stats::dist()},
of course for this parameter to make sense the input cannot be a distance matrix, if it is, then \code{"distfun"}
is ignored).
If the \code{"order"} parameter is \code{NA}, then no seriation is done and the
function prepares the diagram with the user provided ordering
(equivalent to \code{"order"="Identity"}). Alternatively the user may
provide a numeric vector (a permutation of \code{1:number of observations})
according to which the observations will be reordered.

Through the parameter \code{"scale\_data"}, by default \code{TRUE}, the
user has the option to first scale and center the data, via \code{base::scale()}.
The data is next transformed into a distance matrix, using a possibly user provided
distance function.
Then, an ordering of the rows (and columns) of the distance matrix 
is found that optimizes the objective function (under the provided seriation method).
The returned \code{"czek\_matrix"} object resembles the matrix with distances
between the observations. However each entry of this object is not the distance between
the observations but an integer indicating which distance interval 
does the distance belong to. The lower the integer the closer together were the observations.
Importantly in the output \code{"czek\_matrix"} object the original order of the 
rows and columns is retained. The attribute \code{"order"} contains the optimal
ordering as found by the method. It is important to point out that 
the vector in \code{"order"} is the ordering of the rows in the original
data matrix (or rows/columns if the input is a \code{dist} object).
This is particularly important to remember if the input data has row names
that correspond to numbers and the original ordering is not ``\code{1:n}''.

If the user wishes to, then a manual rearrangement, tweaking
of the observations is possible using the provided \code{manual\_reorder} method.
This function also updates the attributes of the \code{"czek\_matrix"} object.
Alternatively, one could make the changes directly in the attribute ``by hand''. 
One then has to also remember
to manually update the criteria attributes, where \code{"criterion\_value"} is
the value of the criterion under which the ordering was optimized, by default
the Hamiltonian path length. Below we give an example of both approaches, 
switching the first two observations. 
\\ \noindent
\code{>library(RMaCzek)}\\ \noindent
\code{>x<-mtcars}\\ \noindent
\code{>czkm<-czek\_matrix(x)}\\ \noindent
\code{>neworder<-attr(czkm,"order")}\\ \noindent
\code{>neworder[1:2]<-neworder[2:1]}\\ \noindent
\code{>czkm\_neworder<-manual\_reorder(czkm,neworder,orig\_data=x)}\\ \noindent
\code{>attr(czkm,"order")<-neworder}\\ \noindent
\code{>attr(czkm,"Um")<-Um\_factor(distMatrix=dist(scale(mtcars)),}\\ \noindent
\code{    order=attr(czkm,"order"),inverse\_um=FALSE)}\\ \noindent
\code{>attr(czkm,"Path\_length")<-attr(czek\_matrix(x,}\\ \noindent
\code{	  ,order=attr(czkm,"order")),"Path\_length")}\\ \noindent
\code{>attr(czkm,"criterion\_value")<-attr(czek\_matrix(x,}\\ \noindent
\code{    order=attr(czkm,"order")),"Path\_length")}\\ \noindent
Apart from a good seriation of the objects a key part is to correctly divide the distances
(or ranks of distances). By default the distances are divided so that each class
has an equal amount of distances (e.g. if \code{"n\_classes"=5}, then each class
will have $20\%$ of the distances). However, the user can easily change this
through the  \code{"interval\_breaks"} parameter. They may provide a vector
summing up to one which will be understood as the fraction of distances
in each class. Otherwise if a vector of positive numbers is provided, 
then these will be the interval breaks of the distances (the vector
has to start with $0$ and end with the largest distance). 
It is important to repeat after \citet{ASolPJas1999} that
different choices lead to emphasizing different aspects 
of the similarities and one has to carefully consider
the choices. As the most useful visualization problem is data specific perception issue
and not an algorithmic one, then it is advisable to try out different 
possibilities and choose the one (or ones) that best seem to capture 
the information in the sample. It might happen, that manual adjustment
possibilities will be crucial.

The function has a number of more advanced options. As already mentioned in Sec. \ref{secCzkDiag}
one can choose to have an asymmetrical diagram---as originally proposed by \citet{JCze1909}.
This is achieved by setting the parameter \code{"original\_diagram"=TRUE} (by default \code{FALSE}).
Then, one can set how to group the ranks of the distances into symbols, using
the parameter \code{"column\_order\_stat\_grouping"}. This grouping is applied to each column.
The default setting is
\citep[same as by][]{JCze1909}, to group first the three most similar to the column
observation, the fourth most similar, the fifth most similar, the sixth most similar
and finally all the remaining (assigning the blank symbol). Here one passes 
a vector with the border ranks, e.g. for \citepos{JCze1909} setting this would 
be \code{c(3,4,5,6)}. In this case the \code{"n\_classes"} parameter is ignored.
The user may also mark a number of objects through the parameter \code{"focal\_obj"},
so that they will not be taking part in the seriation and will be ordered last. This 
could be useful if there are observations that one would want to manually experiment with.

The \code{plot} method for the \code{"czek\_matrix"} class,
plots the \code{"czek\_matrix"} object. The user is able to control
the colours, symbols, sizes of the graphics representing the distances.
Furthermore, the user is given the possibility to manipulate the cell sizes, the axes and labels.
In Czekanowski's original proposal 
the distances could only be represented by different black and white symbols. In our package, for
historical compatibility, we by default provide the distances as black and white symbols 
(but the user is also able to create a grayscale figure or a coloured 
figure, like a heatmap). The entries in the \code{"czek\_matrix"} object correspond to the 
symbol/size of symbol/gray level/colour that is plotted between the two observations. 

The \code{"czek\_matrix"} class also has an associated \code{print} method. If called,
it will print out the ordered objects along with the various factors associated with the
ordering. If the parameter \code{print\_raw} is set to \code{TRUE}, then the actual 
\code{"czek\_matrix"} class object is displayed with its attributes. 
\end{subsection}

\begin{subsection}{Providing a custom seriation method}\label{sbsecSeriation}
Apart from the custom implemented genetic algorithm \pkg{RMaCzek} relies on the 
\pkg{seriation} package to provide methods to order the observations. As the package
allows for user defined seriation methods, the same mechanism can be used to pass
a user defined seriation method to the \code{czek\_matrix()} function. 
Below we show how, the seriation method will
just be a wrapper around \pkg{RMaCzek}'s inbuilt genetic algorithm.
As a reference for this the user may look at the code of the 
\code{RMaCzek:::.register\_seriate\_ga()} function.
The input and output of the seriation function has to conform to the requirements of the 
\pkg{seriation} interface. In particular, based on \code{?seriation::set\_seriation\_method},
the user defined method has to have two formal arguments \code{"x"} the 
distance between observations object and \code{"control"} that is a list with ``additional information
passed on from the function \code{seriation::seriate()}''. The output has to be a list, with
elements in the list corresponding to the dimensions of \code{"x"}.
Each element has to be an object that can be coerced into a 
\code{ser\_permutation\_vector} object, e.g. integer vectors providing the row 
and column ordering.
\\ \noindent
\code{>library(seriation)}\\ \noindent
\code{>library(RMaCzek)}\\ \noindent
\code{>my\_ser\_ga<-function(x,control){RMaCzek:::.seriate\_ga(x,control)}}\\ \noindent
\code{>seriation::set\_seriation\_method(kind="dist",}\\ \noindent
\code{    name="my\_seriate\_ga\_name",definition=my\_ser\_ga,}\\ \noindent
\code{    description="custom seriation method")}\\ \noindent
\code{>x<-mtcars}\\ \noindent
\code{>czek\_matrix(x,order="my\_ser\_ga")}\\ \noindent
Notice that we provide \code{czek\_matrix()} with 
the name of the seriation method and not the defining function object.
\end{subsection}
\end{section}

\begin{section}{Example analyses}\label{secAnalyses}
We begin illustrating the usage of the \pkg{RMaCzek} package by reconsidering
\citepos{JCze1909} original skull data distance matrix. This distance 
matrix is included with \pkg{RMaCzek}, as \code{skulls\_distances}. However we need to emphasize, that
there is an error in \citepos{JCze1909} work. The distance from the 
Neandertal skull to the Galey Hill skull $(10.54)$ does not equal the 
distance from the Galey Hill to the Neandertal skull $(10.504)$. Obviously
this is a minor typographic mistake, and going back to \citep{KSto1908} we believe it
should equal rather $10.504$. However, in the package we wish to include the original
data, hence one should symmetrize the matrix prior to using it. 
\\ \noindent
\code{>sym\_skulls\_distances[5,9]<-10.504} \\ \noindent
Running the code in the script \code{JCze1909.R} found in the Supplementary Material
we obtain Fig. \ref{figJCze1909}. While the ordering is very different we can see that the same
groups are retained. Pithecanthropus and Kannstatt skulls are distinct 
\citep[as][noticed]{JCze1909}. Then, we have the (Spy, Krapina, Neandertal, Gibraltar) 
and (Br\"ux, Galley Hill, Nowosio{\l }ka, Br\"unn, Egisheim) clusters 
\citep[same as][discussed]{JCze1909}. We note that here (unlike in \citepos{JCze1909} presentation)
Pithecanthropus is before the ``Neandertal'' cluster---consistent with the timeline ordering proposed by 
\citet{GSch1906}. The Nowosi{\'o}{\l }ka skull is firmly placed within
the ``H. sapiens'' cluster---just as \citet{JCze1909} wrote.
Hence, while the same general groups are found as \citet{JCze1909} presented
them, inside each the observations are reordered. 

\begin{figure}[p]
\begin{center}
\includegraphics[width=0.32\textwidth]{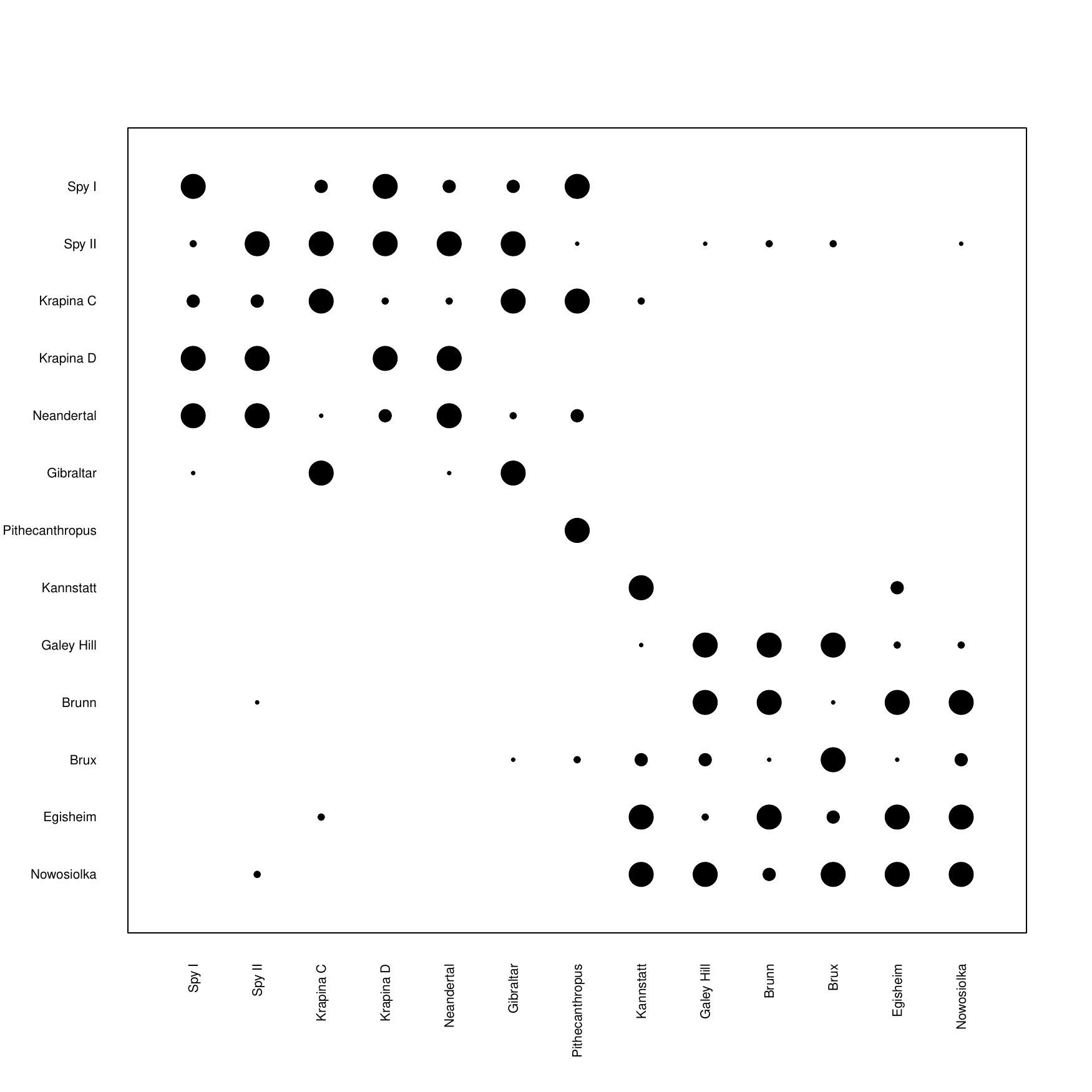}
\includegraphics[width=0.32\textwidth]{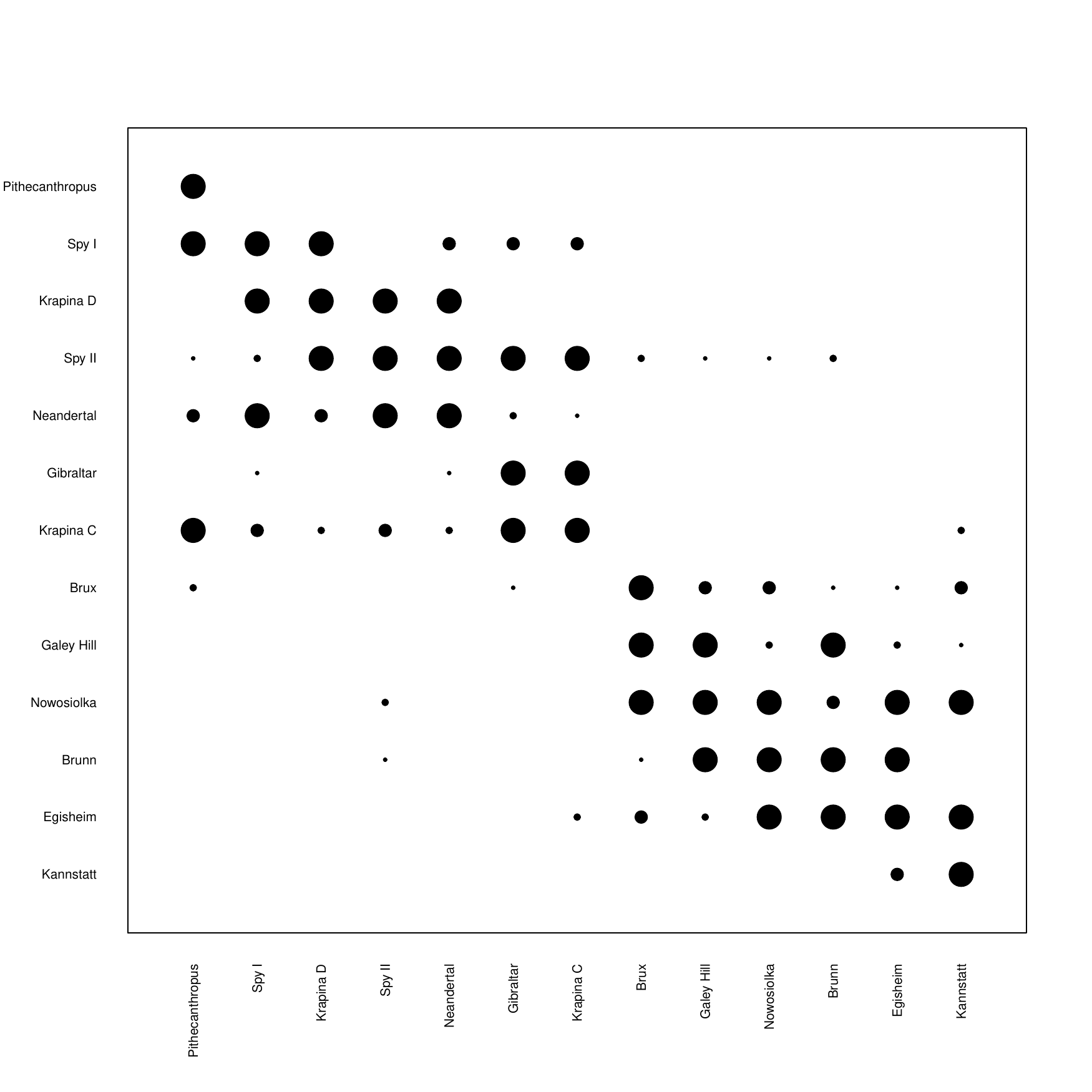}
\includegraphics[width=0.32\textwidth]{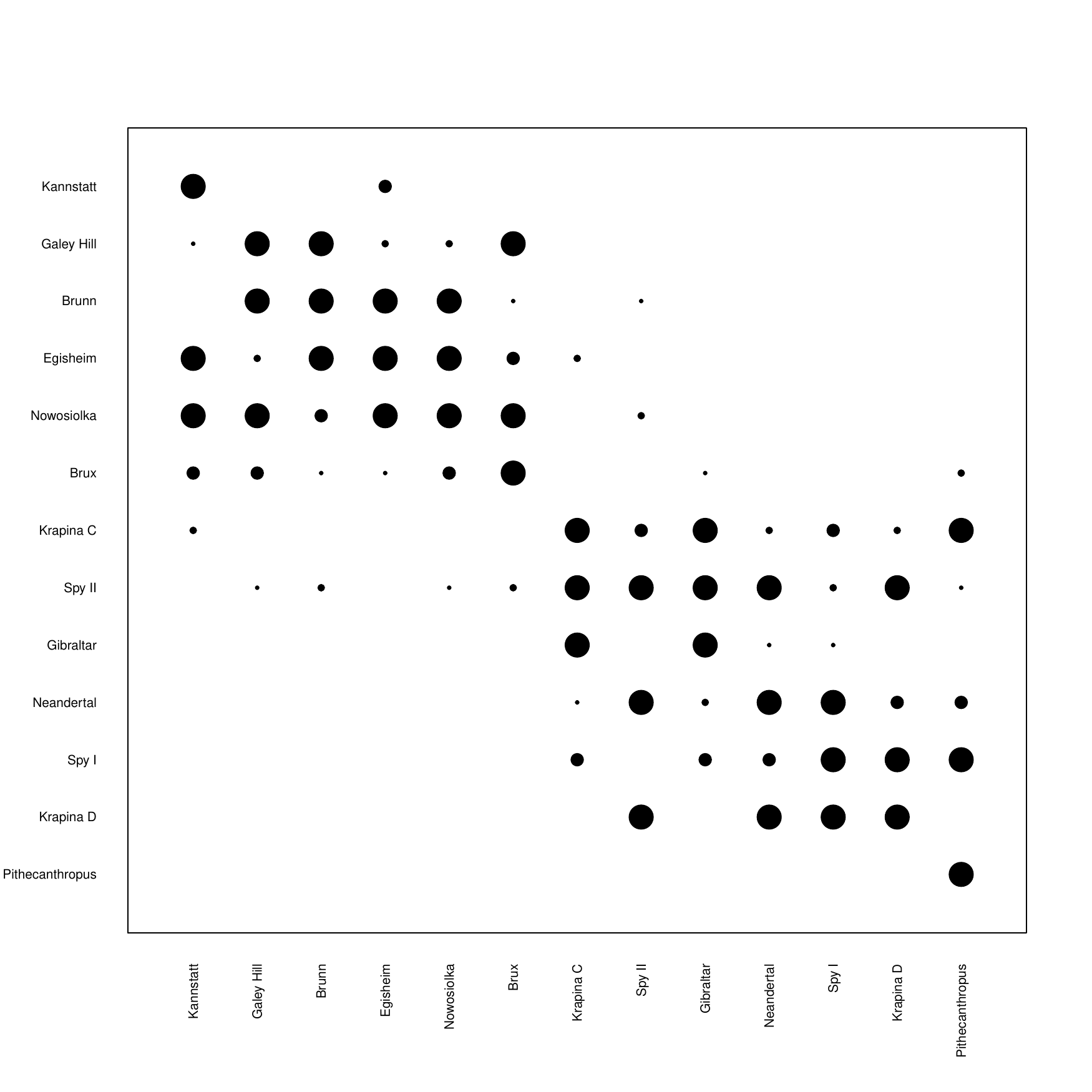} \\
\includegraphics[width=0.32\textwidth]{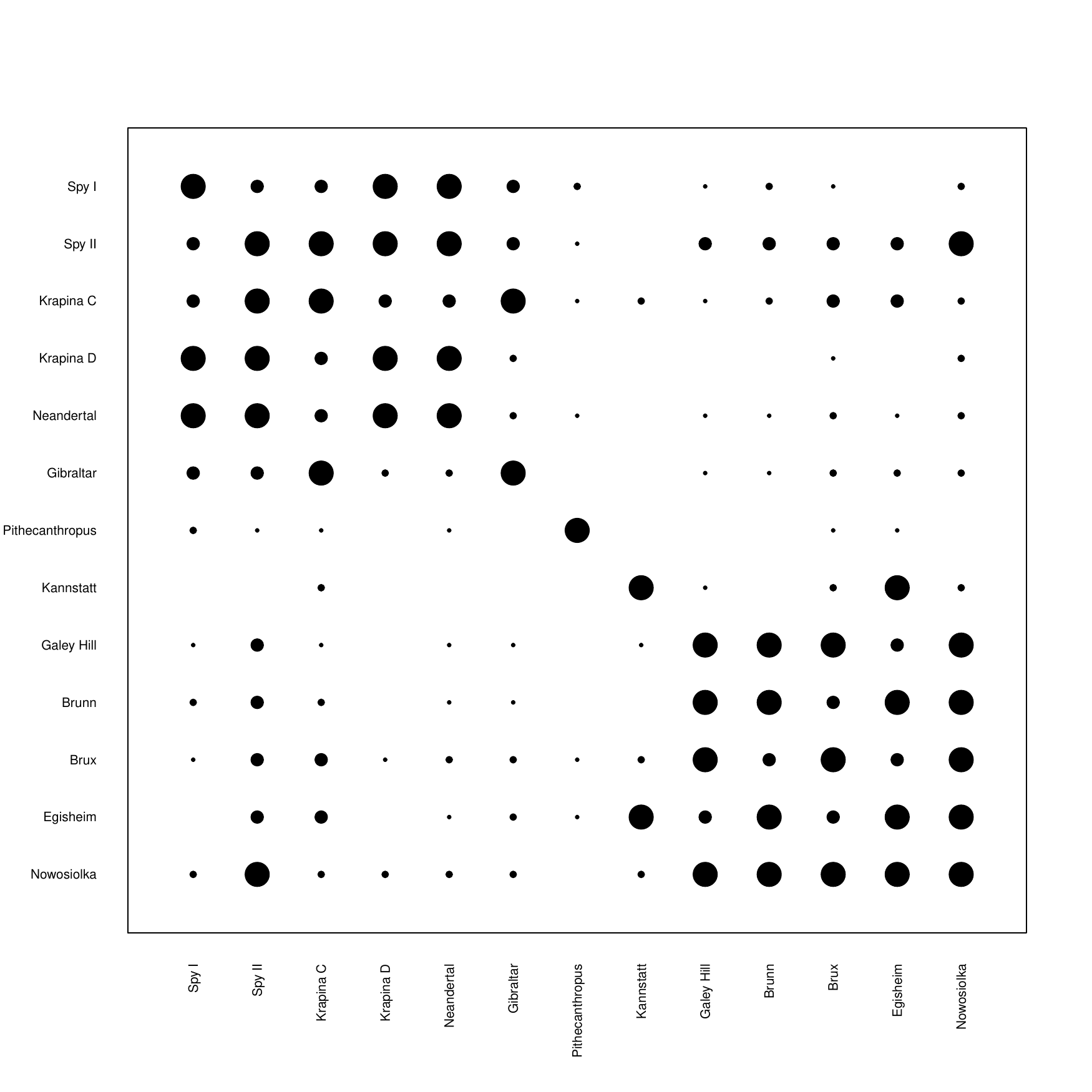} 
\includegraphics[width=0.32\textwidth]{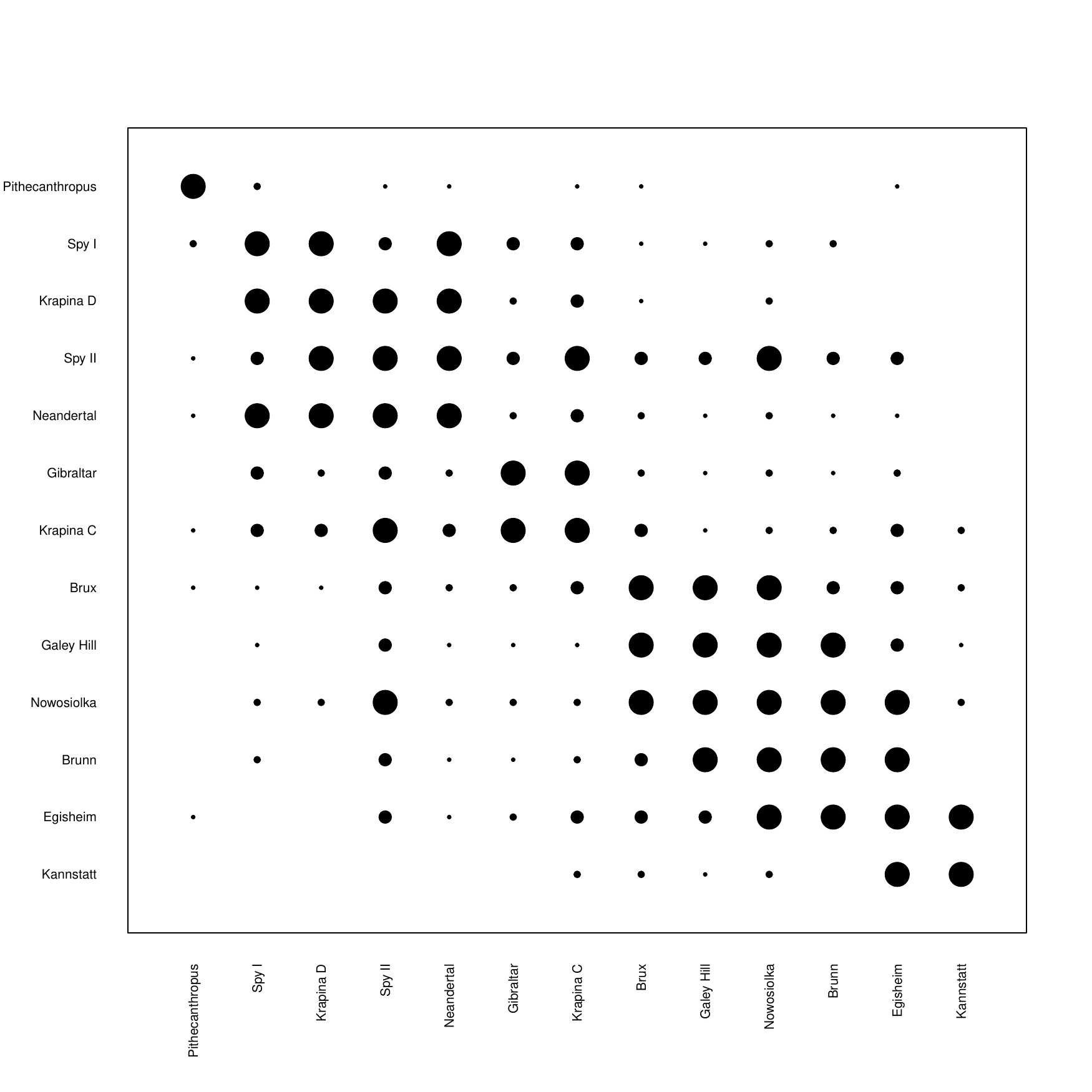} 
\includegraphics[width=0.32\textwidth]{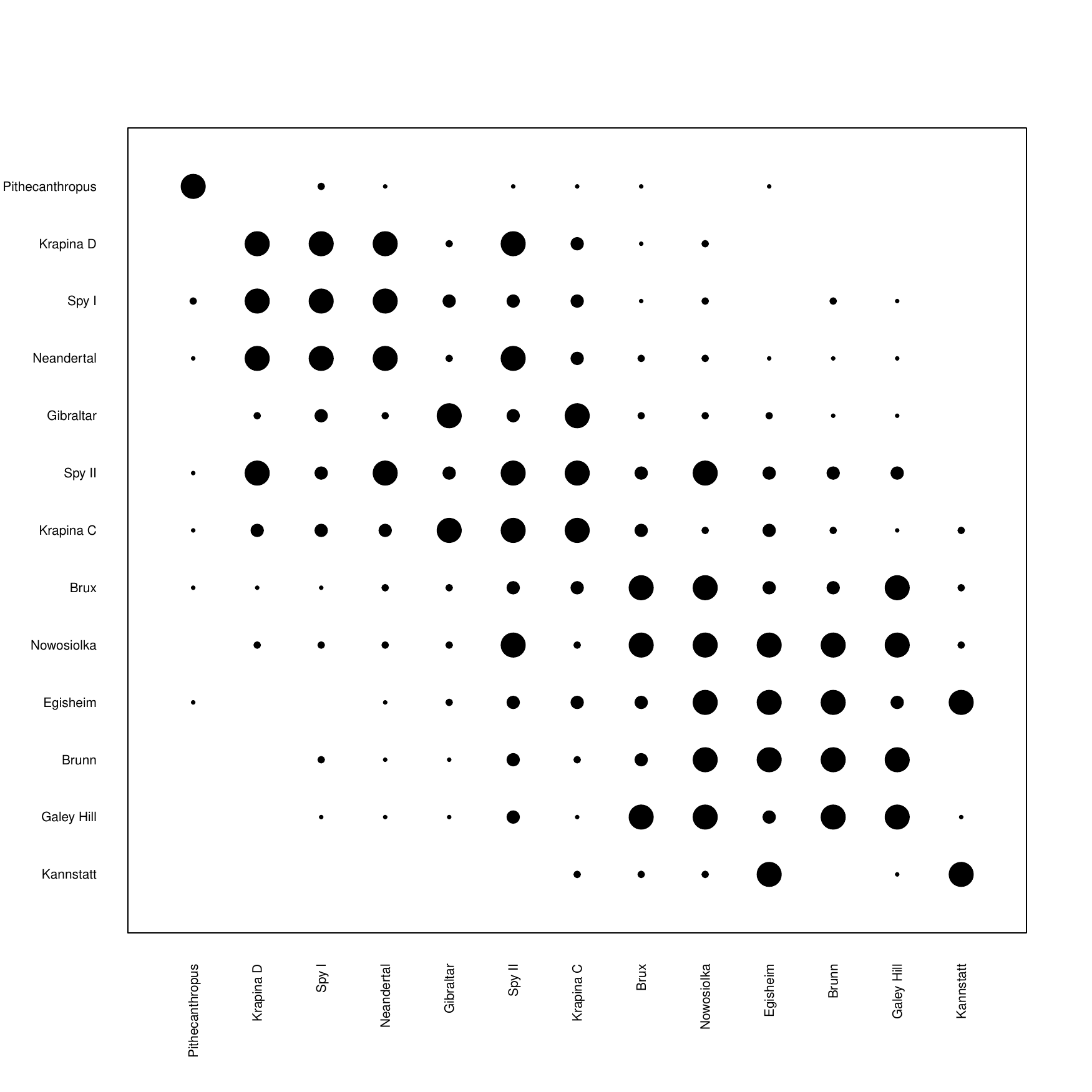}
\end{center}
\caption{
Visualization of \citepos{JCze1909} original distance matrix between
archaic humans' skulls. Top row: diagrams with \citepos{JCze1909}
original proposal presenting in each column the most similar rows,
bottom row: symmetric diagram directly visualizing the distance matrix.
Left column: data presented in \citepos{JCze1909} original ordering
($U_{m} = 3.133$, path length = $85.114$), center column:
diagram produced by \texttt{order="OLO"}
($U_{m} = 2.56$, path length = $59.953$) and
right column:
diagram produced by \texttt{order="QAP\_2SUM"}
($U_{m} = 2.483$, path length = $75.928$).
\label{figJCze1909} 
}
\end{figure}

We next look at \citepos{ASolPJas1999} urns dataset 
(included as \code{urns} in \pkg{RMaCzek})
that was used to illustrate the original
\pkg{MaCzek} program. \citet{ASolPJas1999} considered nine measurements 
(height (WYS), rim diameter (SW), maximal diameter (MWB), bottom diameter (SD), average wall thickness (GS), 
average bottom thickness (GD) and three indices describing proportions of the vessel (W--A, W--B, W--D)
of fourteen urns from cremation graves  excavated at
Paprotki Kolonia $12$ in Poland. One urn, ``gr. 52--1'', was removed from 
further analyses as it contained missing values on seven out of nine variables. 

After running the \code{ASolPJas1999.R} script from the Supplementary Material,
we obtain the ordering presented in Fig. \ref{figASolPJas1999}. We can see that
the ordering found by \code{order="OLO"} captures the three main clusters 
(and the same order between them)
of \citepos{ASolPJas1999} found ordering:
$\{\{$gr. 2--1, gr. 5B--1, gr. 8--1, gr. 27--1, gr. 3--1, gr. 6A--4$\},
\{$gr. 46--2, gr. 66--2, gr. 64--2, gr. 72--2, gr. 23--?$\},
\{$gr. 58--3, gr. 113--1, gr. 102--2$\}\}$ while
``QAP\_2SUM'' finds only the two main clusters
$\{\{$gr. 2--1, gr. 5B--1, gr. 8--1, gr. 27--1, gr. 3--1, gr. 6A--4$\},
\{$gr. 46--2, gr. 66--2, gr. 64--2, gr. 72--2, gr. 23--?,
gr. 58--3, gr. 113--1, gr. 102--2$\}\}$.
Despite finding the same main structure it seems that the \code{"OLO"} method
better arranges inside the clusters. This is actually evident from the ``asymmetric
diagrams'', where the most similar are marked in each column. The \code{"OLO"} one
is much more ``diagonal'' than \citepos{ASolPJas1999} original one
or the \code{"QAP\_2SUM"} one.

\begin{figure}[p]
\begin{center}
\includegraphics[width=0.32\textwidth]{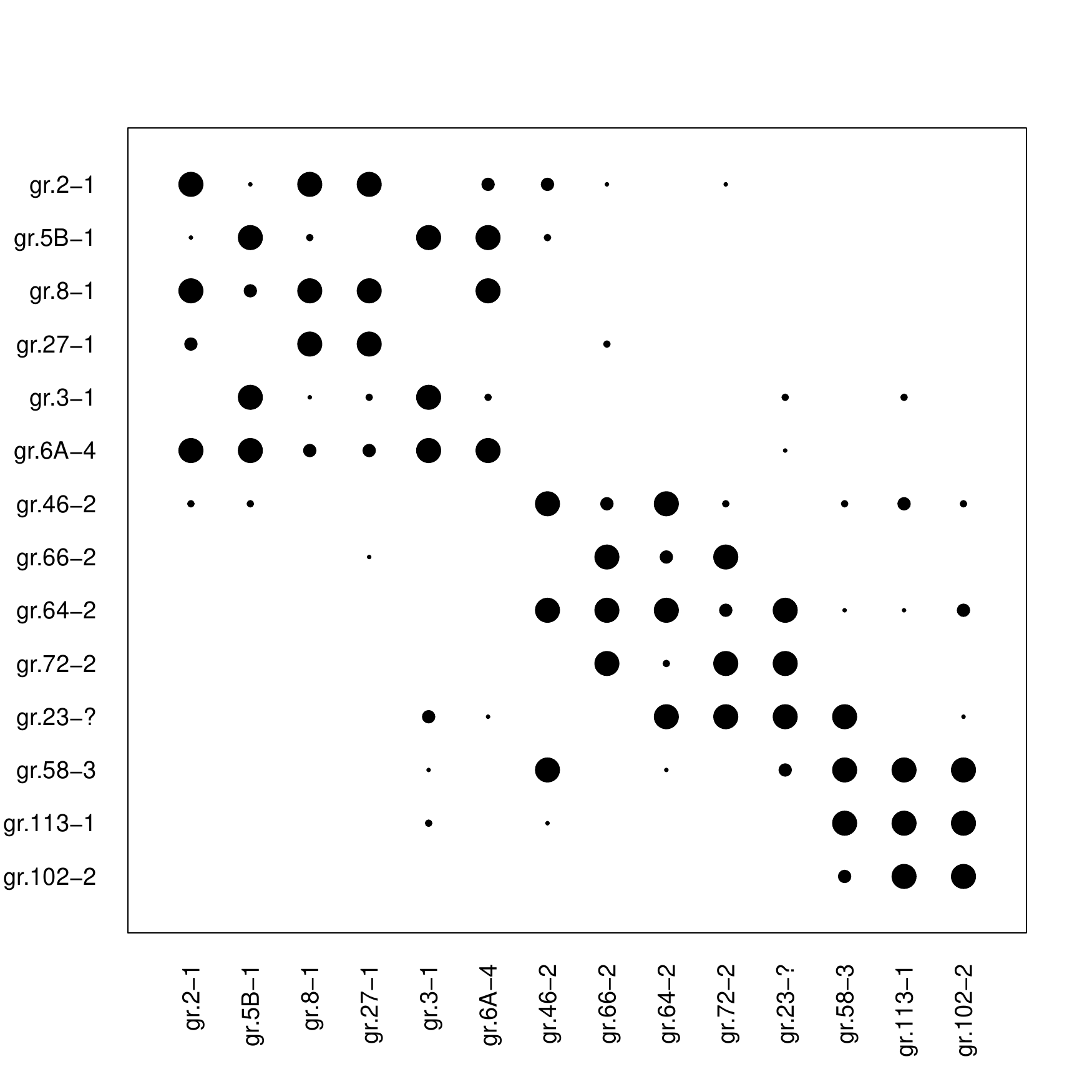}
\includegraphics[width=0.32\textwidth]{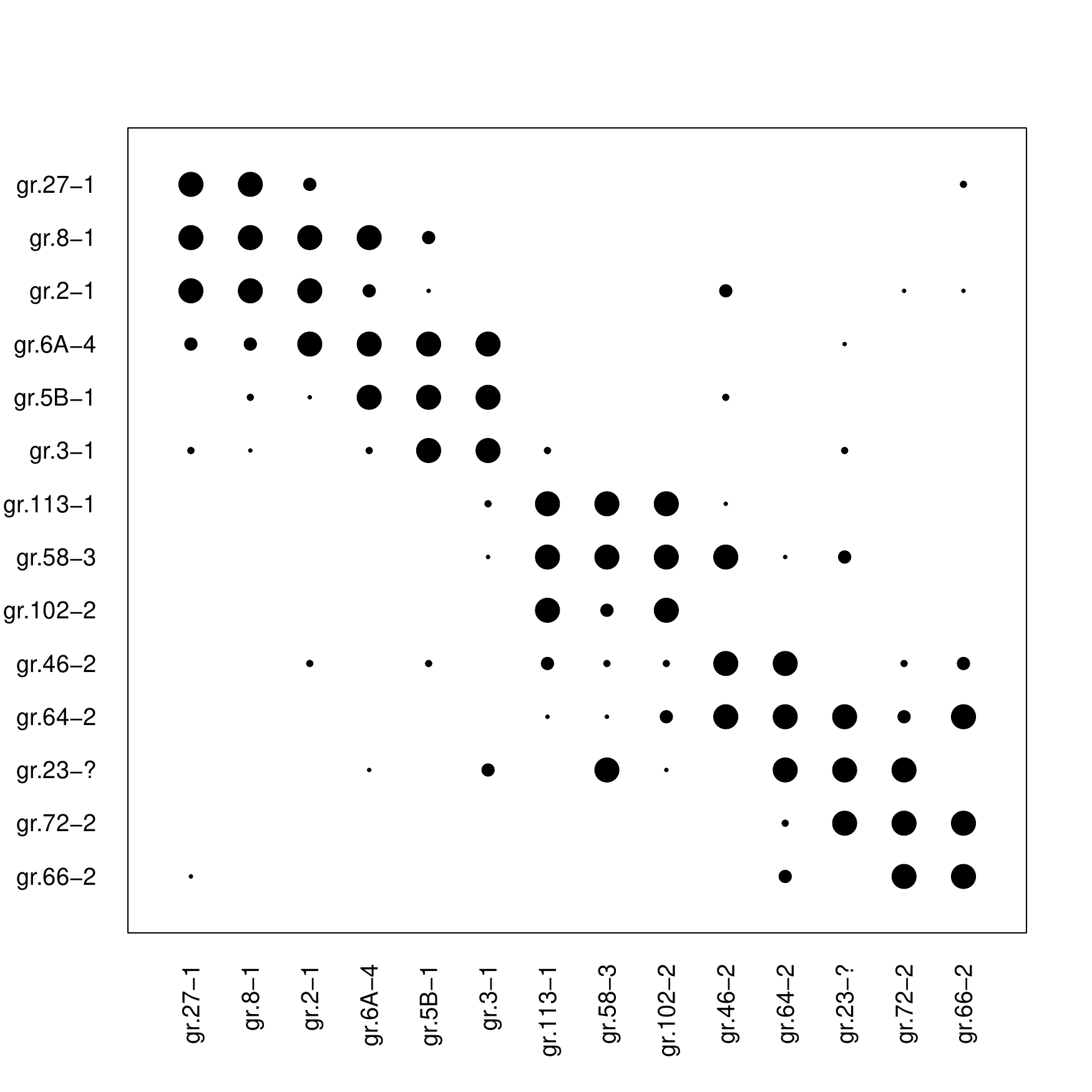}
\includegraphics[width=0.32\textwidth]{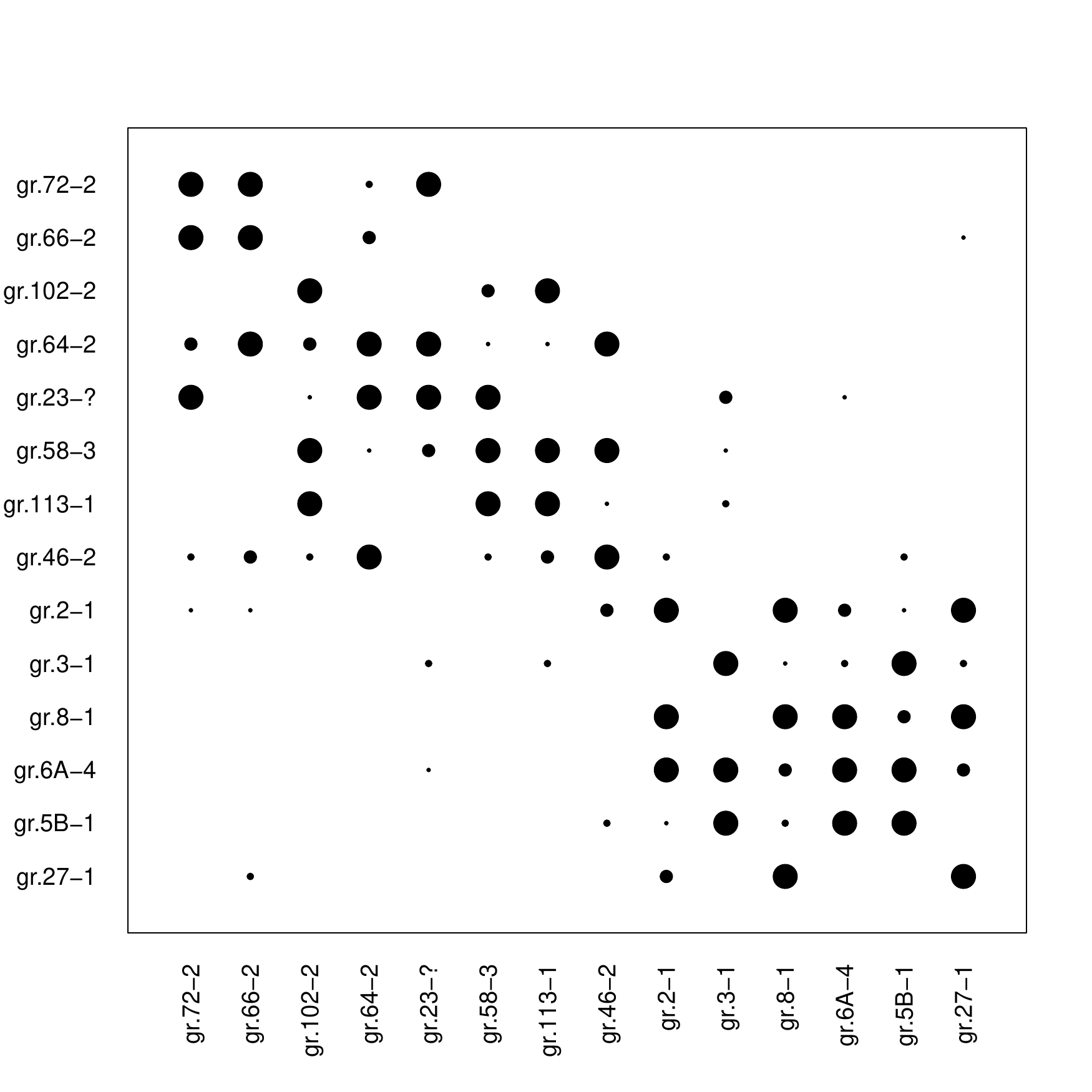} \\
\includegraphics[width=0.32\textwidth]{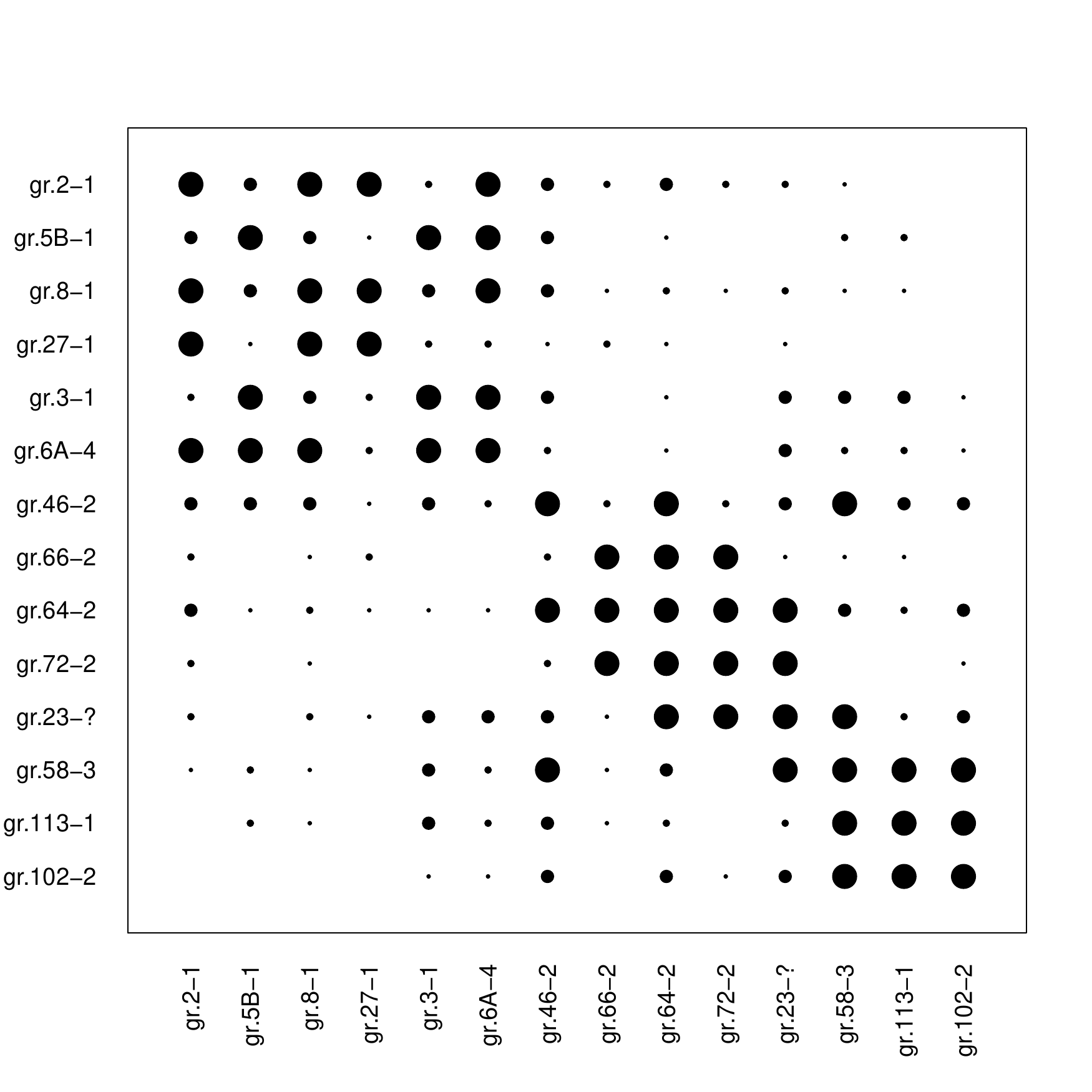} 
\includegraphics[width=0.32\textwidth]{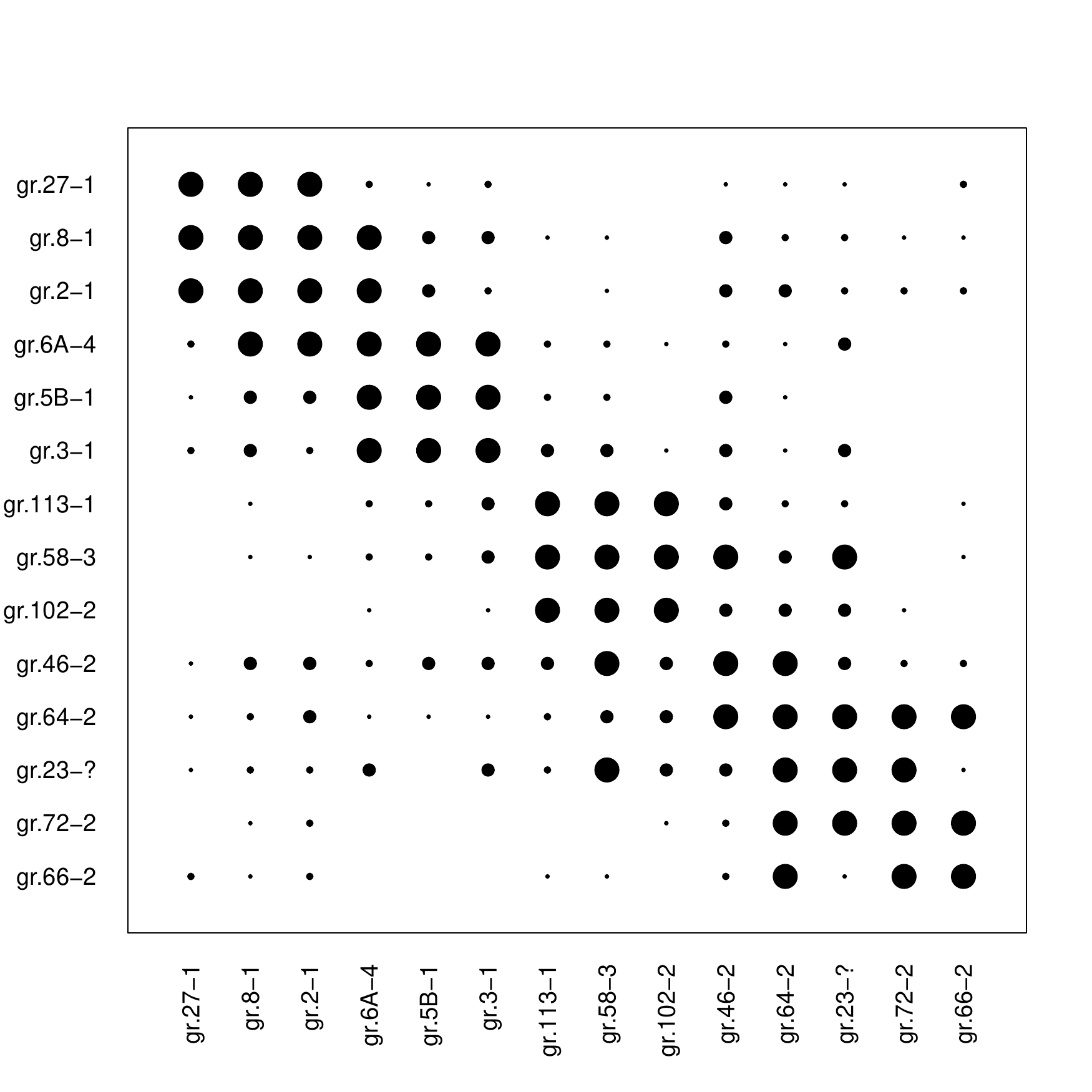} 
\includegraphics[width=0.32\textwidth]{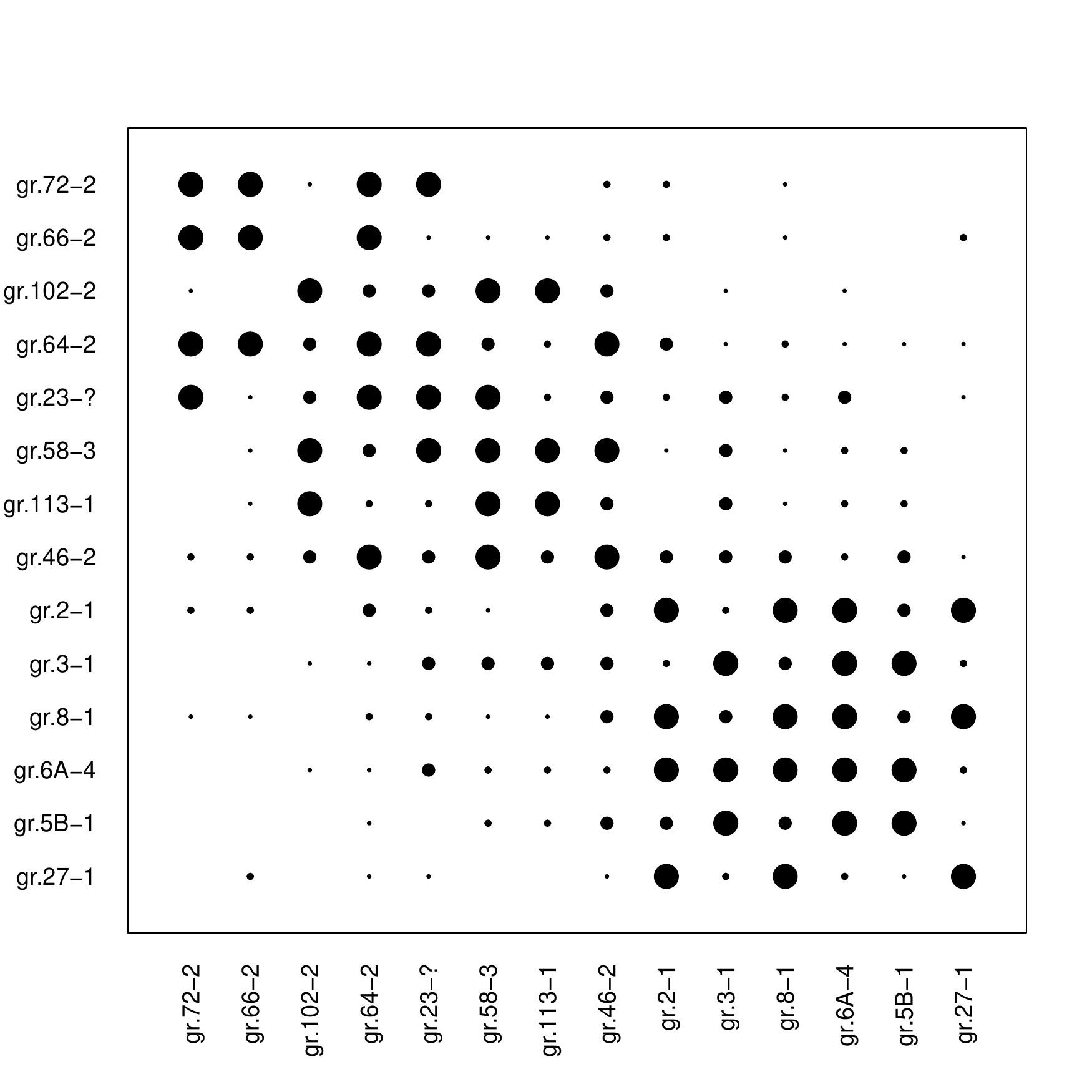}
\end{center}
\caption{
Visualization of \citepos{ASolPJas1999} urns data. 
Top row: diagrams with \citepos{JCze1909}
original proposal presenting in each column the most similar rows,
bottom row: symmetric diagram directly visualizing the distance matrix.
Left column: data presented by \citet{ASolPJas1999} found by \pkg{MaCzek} ordering
($U_{m} = 5.933$, path length = $36.111$, $U_{m}$ was minimized), center column:
diagram produced by \texttt{order="OLO"}
($U_{m} = 5.873$, path length = $28.515$, path length was minimized) and
right column:
diagram produced by \texttt{order="QAP\_2SUM"}
($U_{m} =5.69$, path length = $38.772$, $U_{m}$ was minimized).
\label{figASolPJas1999} 
}
\end{figure}

As a third example we consider \citepos{ASol2000} Akkadian cylinder seals depicting
the Serpent God dataset 
(included in the \pkg{RMaCzek} package as \code{seals\_similarities}). 
It is a similarity matrix between $37$ seals with $100$
as maximum similarity. The original dataset consisted of $39$ seals
but two ($36$ and $39$) had to be removed ``due to a lack of data in the literature''
\citep{ASol2000}. Each seal is described by a binary vector of $22$ variables,
each variable is the presence or absence (or missing data due to damage of the seal)
of some attribute. Examples of the attributes are e.g. ``Serpent God is on the right'',
``end of the Serpent God's tail is clear and raised'', ``the Serpent God holds a goblet''.
Before passing the data to the \code{RMaCzek::czek\_matrix()} function one has to 
set $100$ on the diagonal (\code{NA} in the original data) and change it 
to a distance matrix as \code{100-similarity},
\\ \noindent
\code{>diag(seals\_similarities)<-100} \\ \noindent
\code{>seals\_distances<-as.dist(100-seals\_similarities)} \\ \noindent
In order to make the symmetric graph
as similar as possible to \citepos{ASol2000} we need to set the number of classes and 
interval breaks appropriately.
We found \code{n\_classes=4,interval\_breaks=c(0,40,60,80,100)}.
to be the best.

\begin{figure}[p]
\begin{center}
\includegraphics[width=0.32\textwidth]{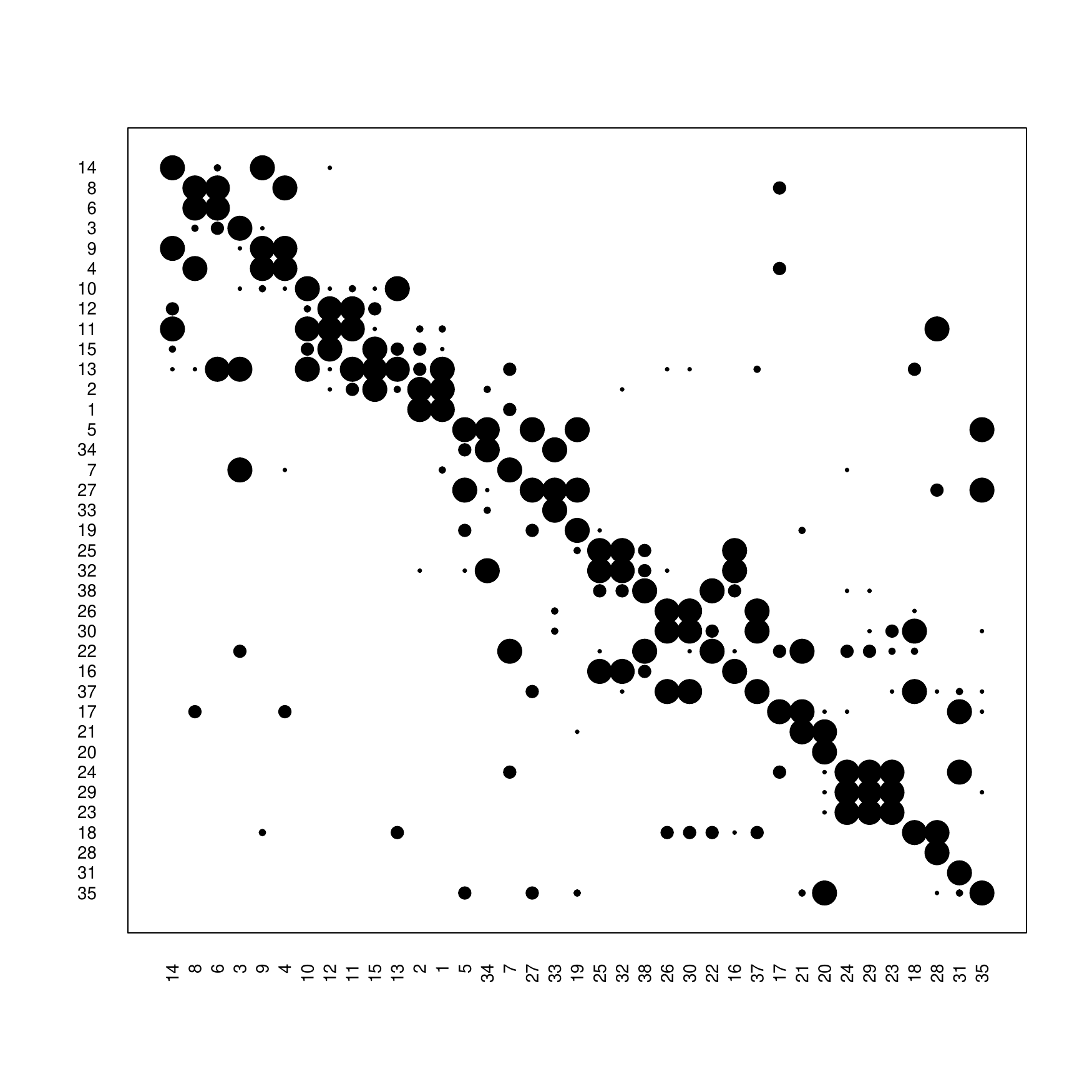}
\includegraphics[width=0.32\textwidth]{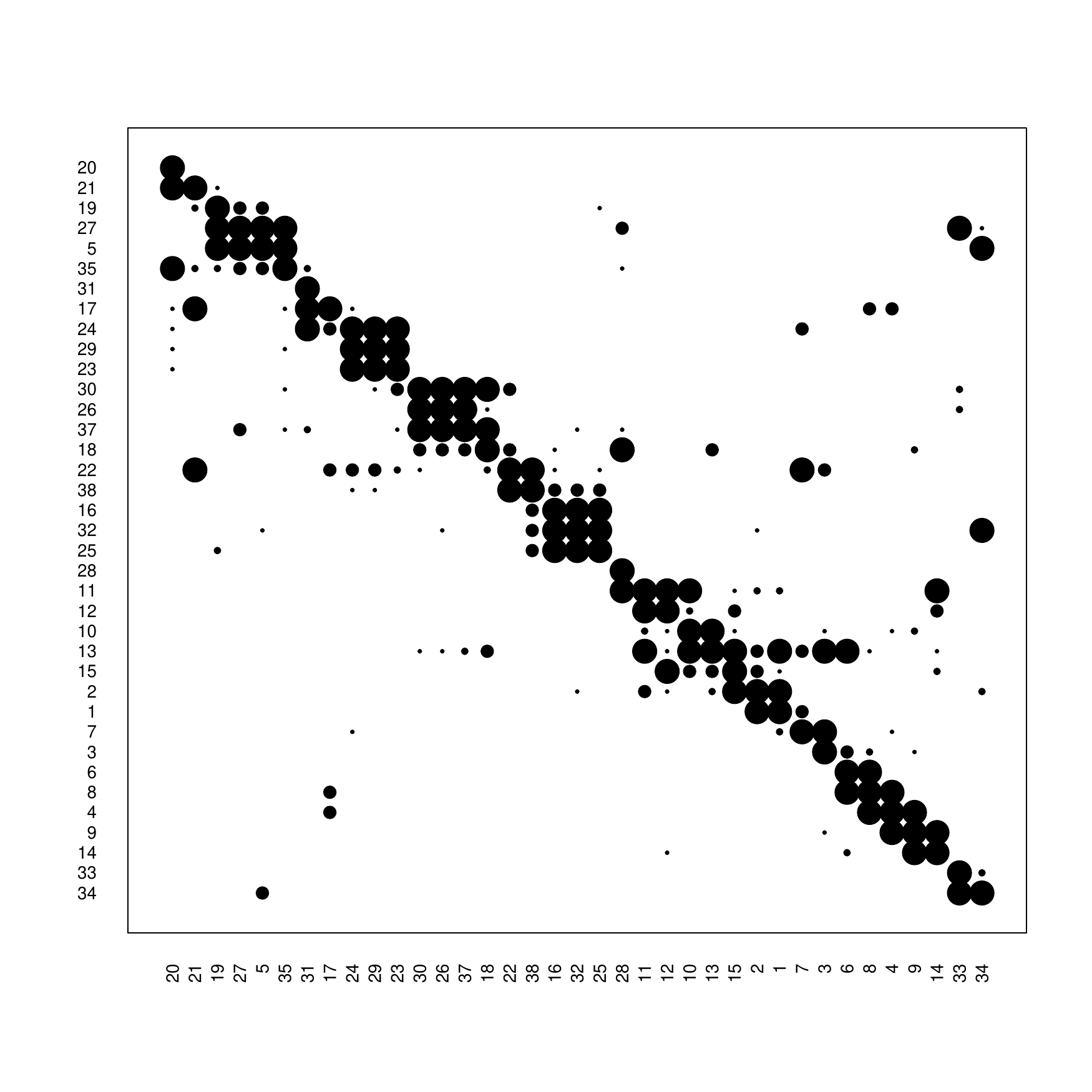}
\includegraphics[width=0.32\textwidth]{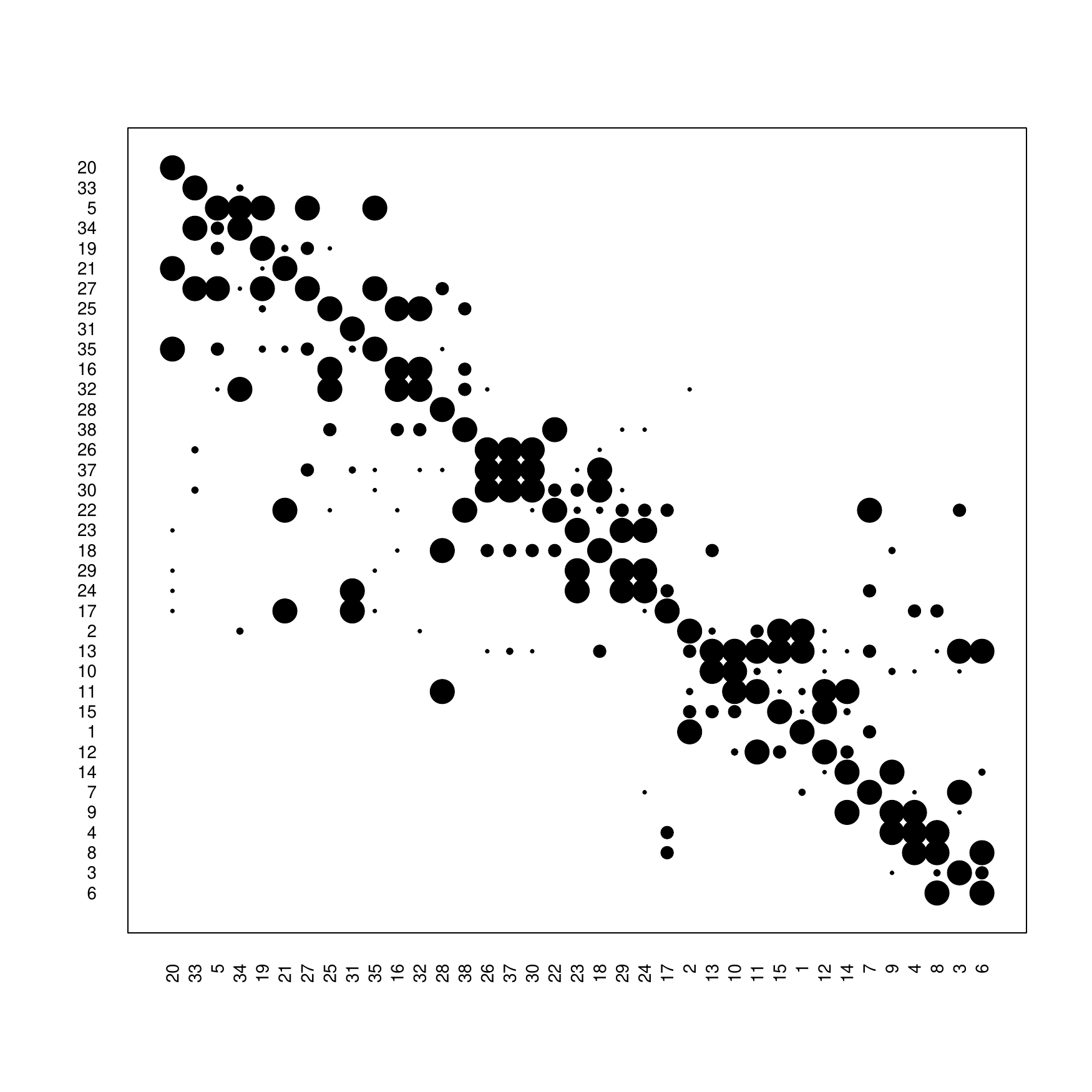} \\
\includegraphics[width=0.32\textwidth]{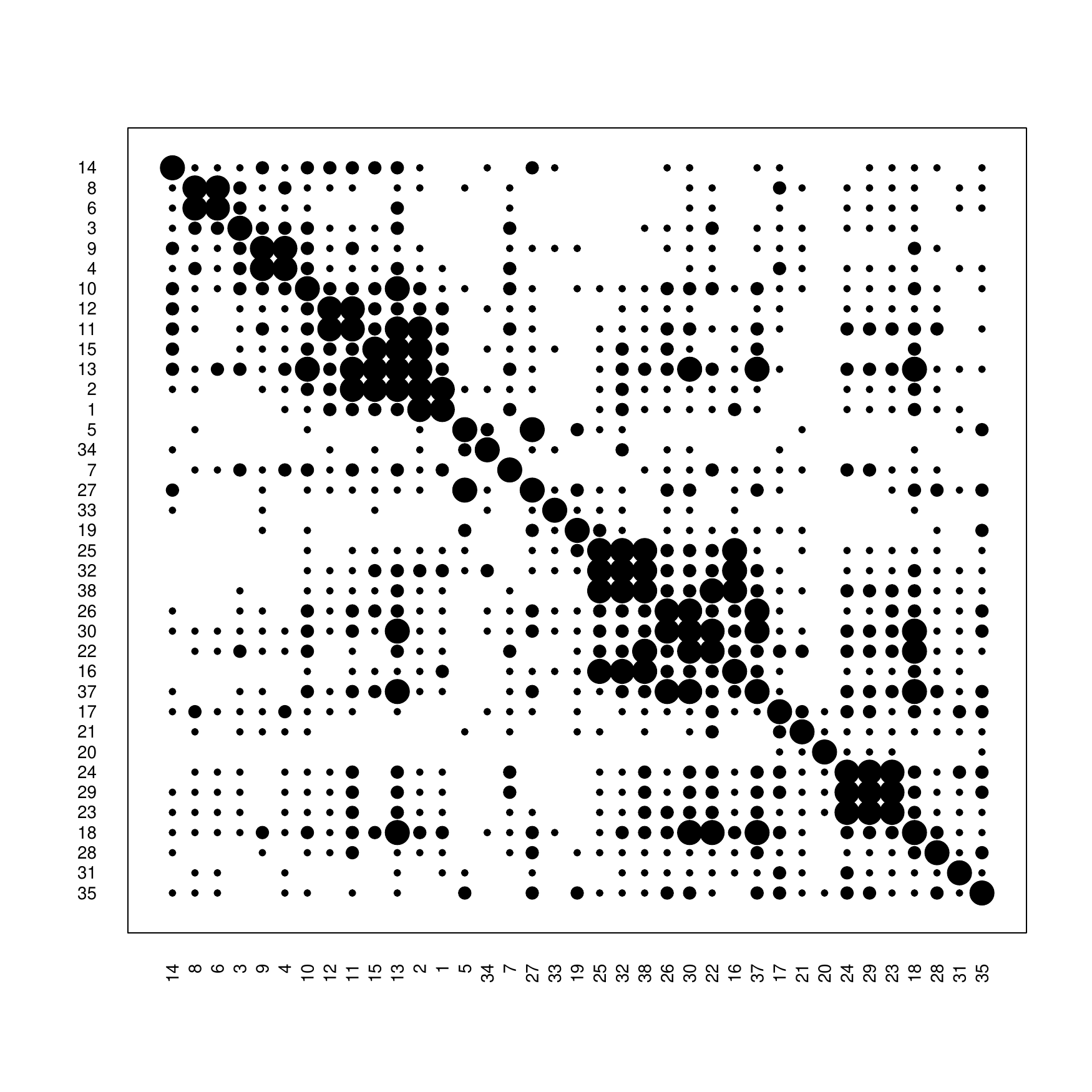} 
\includegraphics[width=0.32\textwidth]{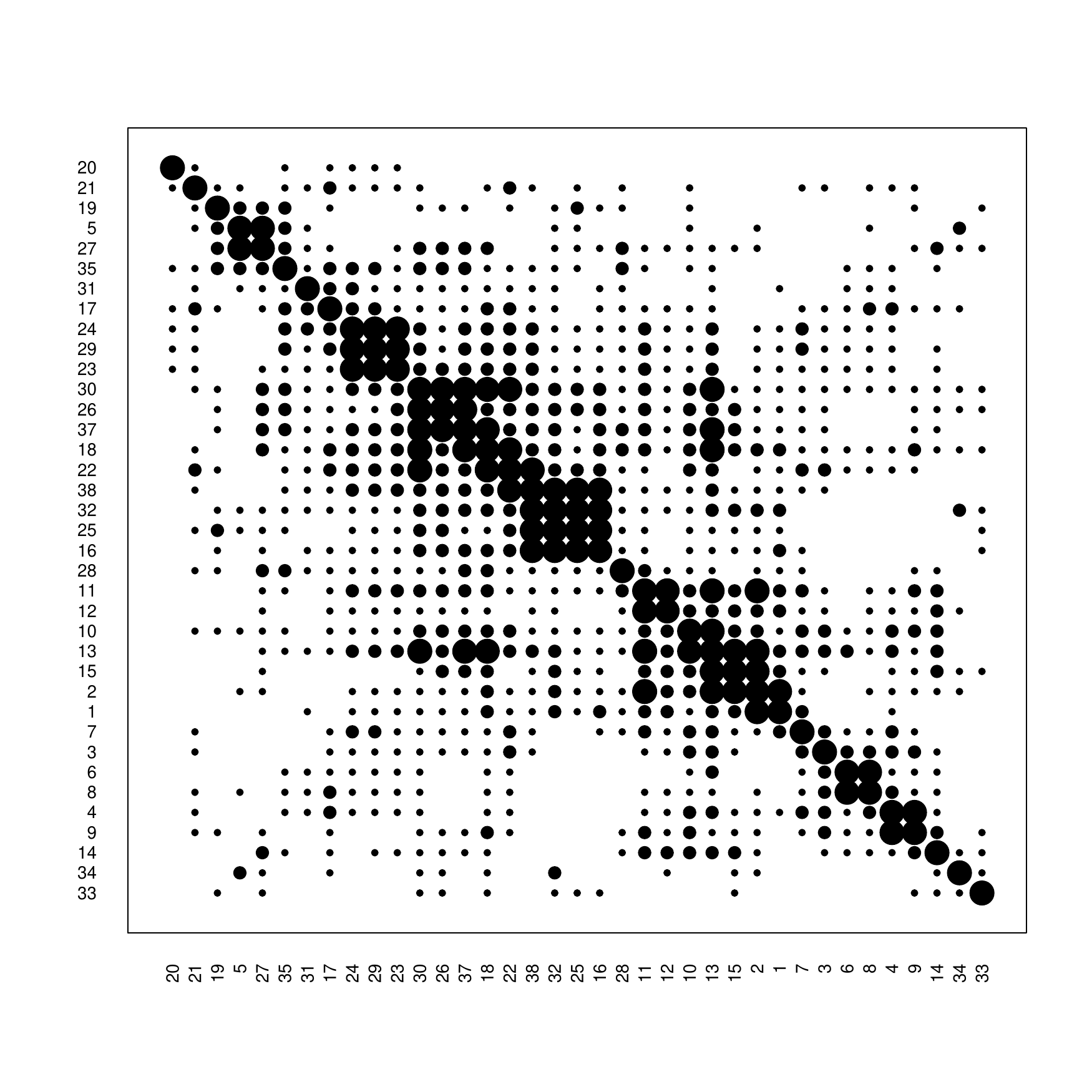} 
\includegraphics[width=0.32\textwidth]{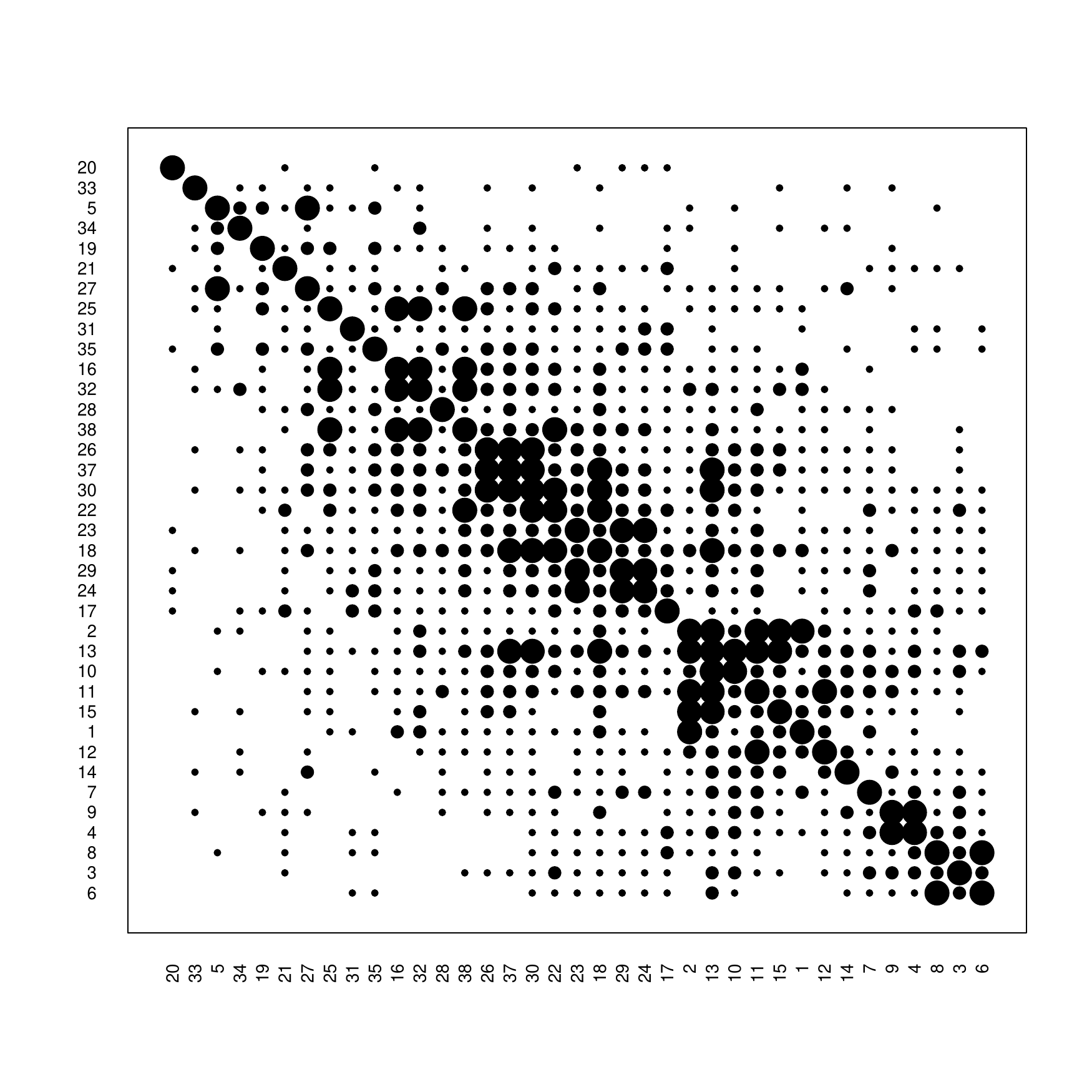}
\end{center}
\caption{
Visualization of \citepos{ASol2000} seals data. 
Top row: diagrams with \citepos{JCze1909}
original proposal presenting in each column the most similar rows,
bottom row: symmetric diagram directly visualizing the distance matrix.
Left column: data presented by \citet{ASol2000} found by \pkg{MaCzek} ordering
($U_{m} = 3.029$, path length = $1892$, $U_{m}$ was minimized), center column:
diagram produced by \texttt{order="OLO"}
($U_{m} = 2.867$ (top); $U_{m} = 2.862$ (bottom), path length = $1490$, path length was minimized,
two orderings with identical path length were found) and
right column:
diagram produced by \texttt{order="QAP\_2SUM"}
($U_{m} =2.779$, path length = $2035$, $U_{m}$ was minimized).
\label{figASol2000} 
}
\end{figure}

After running the script \code{ASol2000.R} from the Supplementary Material
we can see in Fig. \ref{figASol2000} \citepos{ASol2000} original
arrangement from the \pkg{MaCzek} program and \pkg{RMaCzek}'s
\code{"OLO"} and \code{"QAP\_2SUM"} arrangements. All are strongly diagonal,
consistent with \citepos{ASol2000} findings. Also it seems that
the \code{"OLO"} method captured the main clusters that \citet{ASol2000}
did, e.g. $\{\{11,12,10,13,15,02,01\},\{24,29,23\}\}$.
Also, \code{"OLO"} groups together
$\{16,22,25,32,38 \}$, similarly as \citet{ASol2000} did.
Interestingly, in the seals dataset  there are two
arrangements that have the same path length. They differ
by the relative arrangement of the observations 
$\{5,27\}$, $\{16,25,32\}$ and $\{33,34\}$. By the $U_{m}$ factor it would seem that
$\{5,27\}$, $\{32,25,16\}$ and $\{34,33\}$  (symmetric diagram) is the better rearrangement, instead of
$\{27,5\}$, $\{16,32,25\}$ and $\{33,34\}$ (asymmetric diagram). The \code{"QAP\_2SUM"} method did seem to provide 
visually slightly less appealing  diagrams---especially the symmetric one. In this case,
the best similarities seem to be more scattered around the diagonal, than in the 
\code{"OLO"} and \citepos{ASol2000} original orderings. However, one also captures the clusters
$\{02,13,19,11,15,01,12\}$ and $\{29,24\}$.

Finally, we turn to re-analyzing a dataset from a different field---socio--economics.
It is a study of Internet availability for pupils at schools in $36$ counties of 
the Silesia Voivodeship, Poland \citep{KWar2015}. Each county is characterized by
five variables---number of students per one computer with Internet access in
upper secondary schools (liceum),
number of students per one computer with Internet access in
secondary schools (gimnazjum),
number of students per one computer with Internet access in
primary schools, share of upper secondary schools with computers with
Internet access available for students and  
share of secondary schools with computers with
Internet access available for students. 
\citet{KWar2015} normalized (mean centred, scaled by standard deviation) the 
measurements and then calculated the Euclidean distance between each pair of counties.
The distance matrix between the counties is included in \pkg{RMaCzek} as \code{internet\_availability}. 
In Fig. \ref{figKWar2015} we can see her ordering of the data, alongside \pkg{RMaCzek}'s ordering,
according to the script \code{KWar2015.R} in the Supplementary Material.

\begin{figure}[p]
\begin{center}
\includegraphics[width=0.32\textwidth]{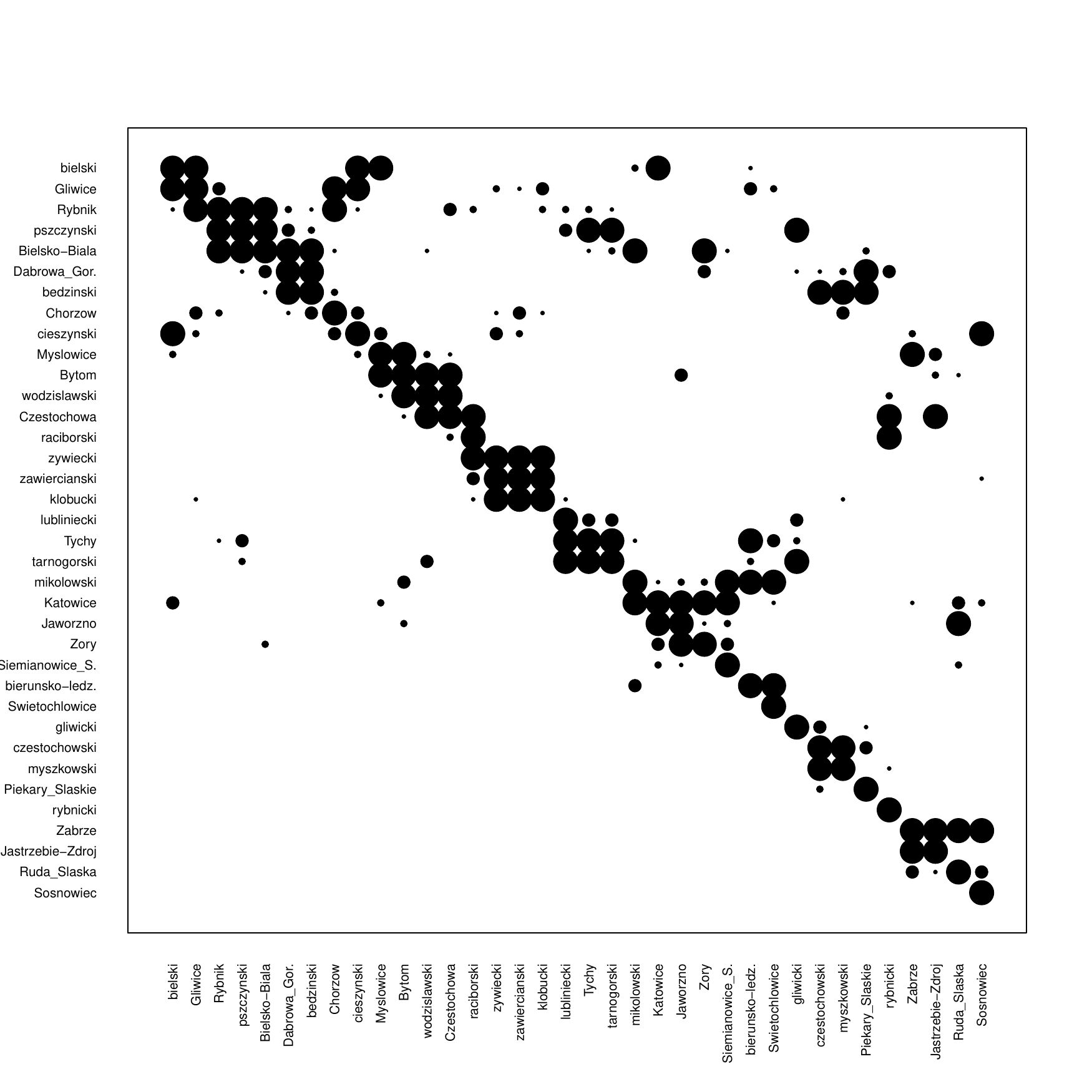}
\includegraphics[width=0.32\textwidth]{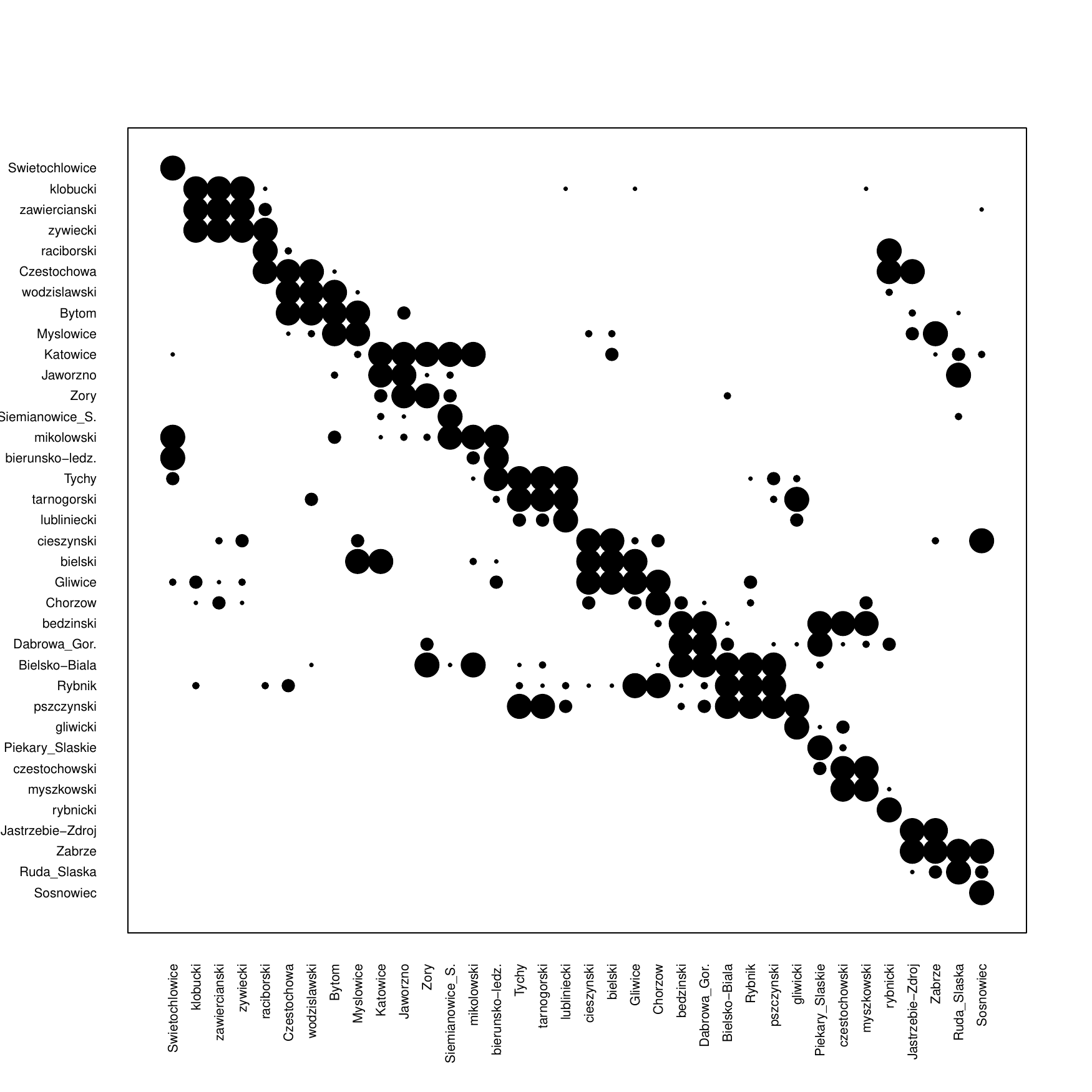}
\includegraphics[width=0.32\textwidth]{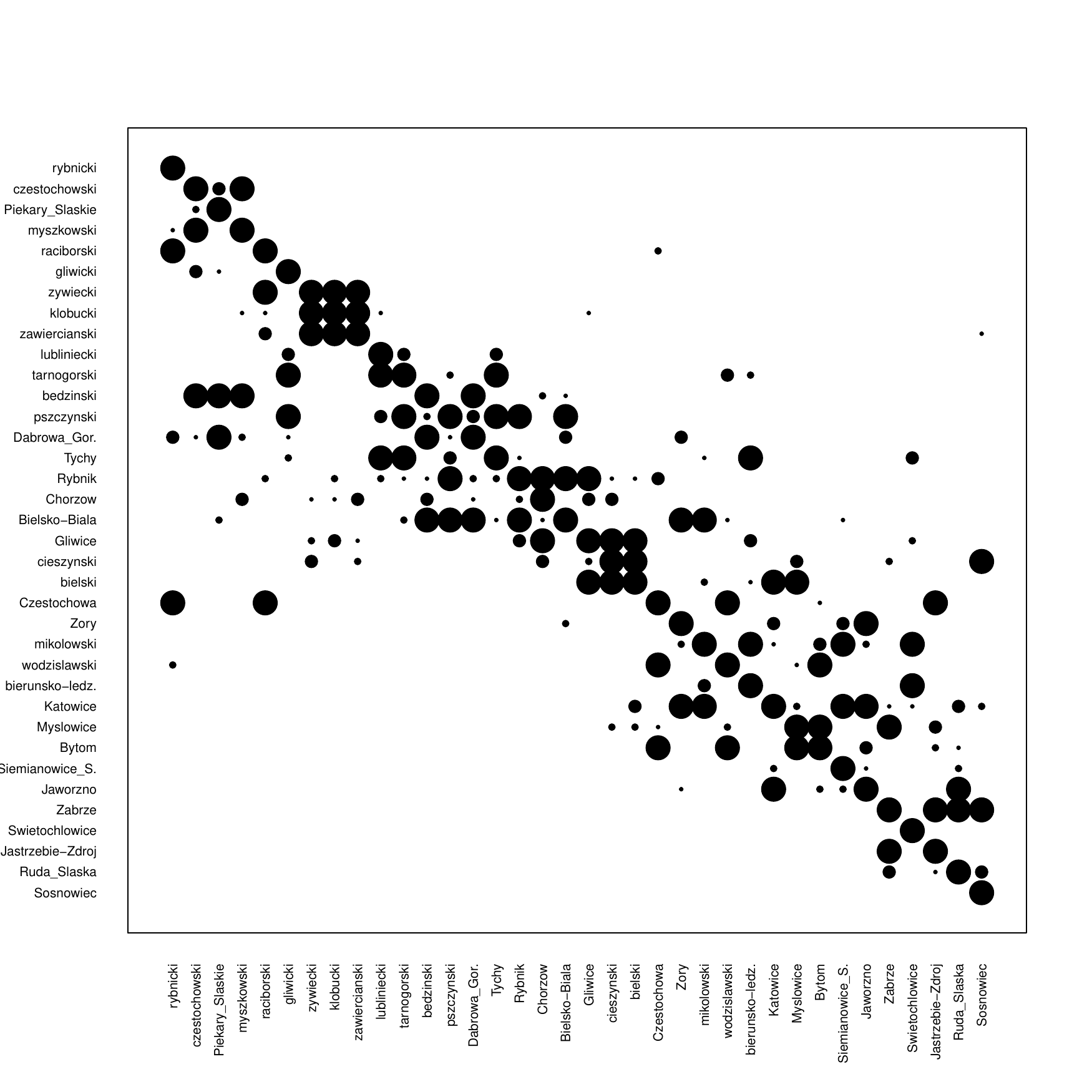} \\
\includegraphics[width=0.32\textwidth]{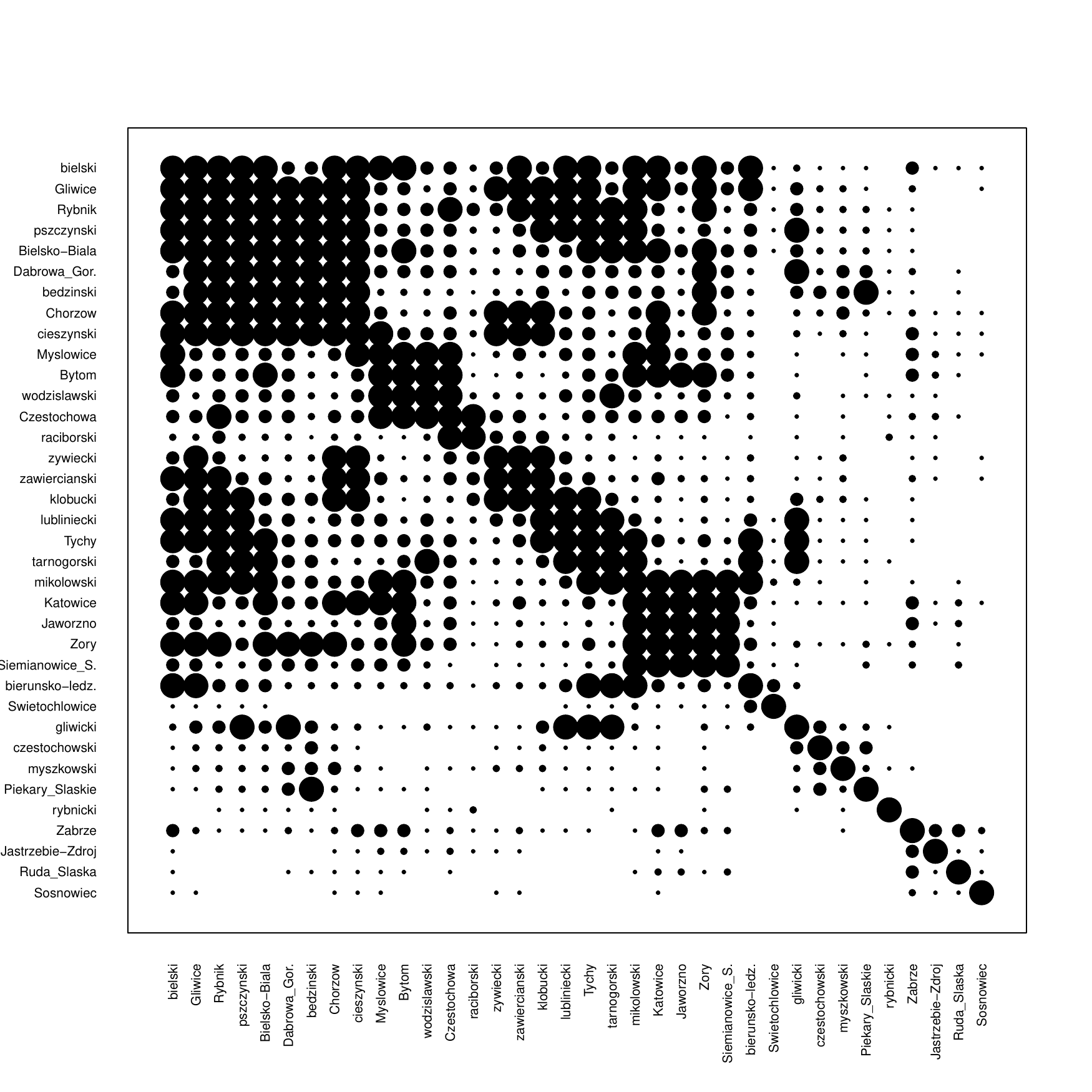} 
\includegraphics[width=0.32\textwidth]{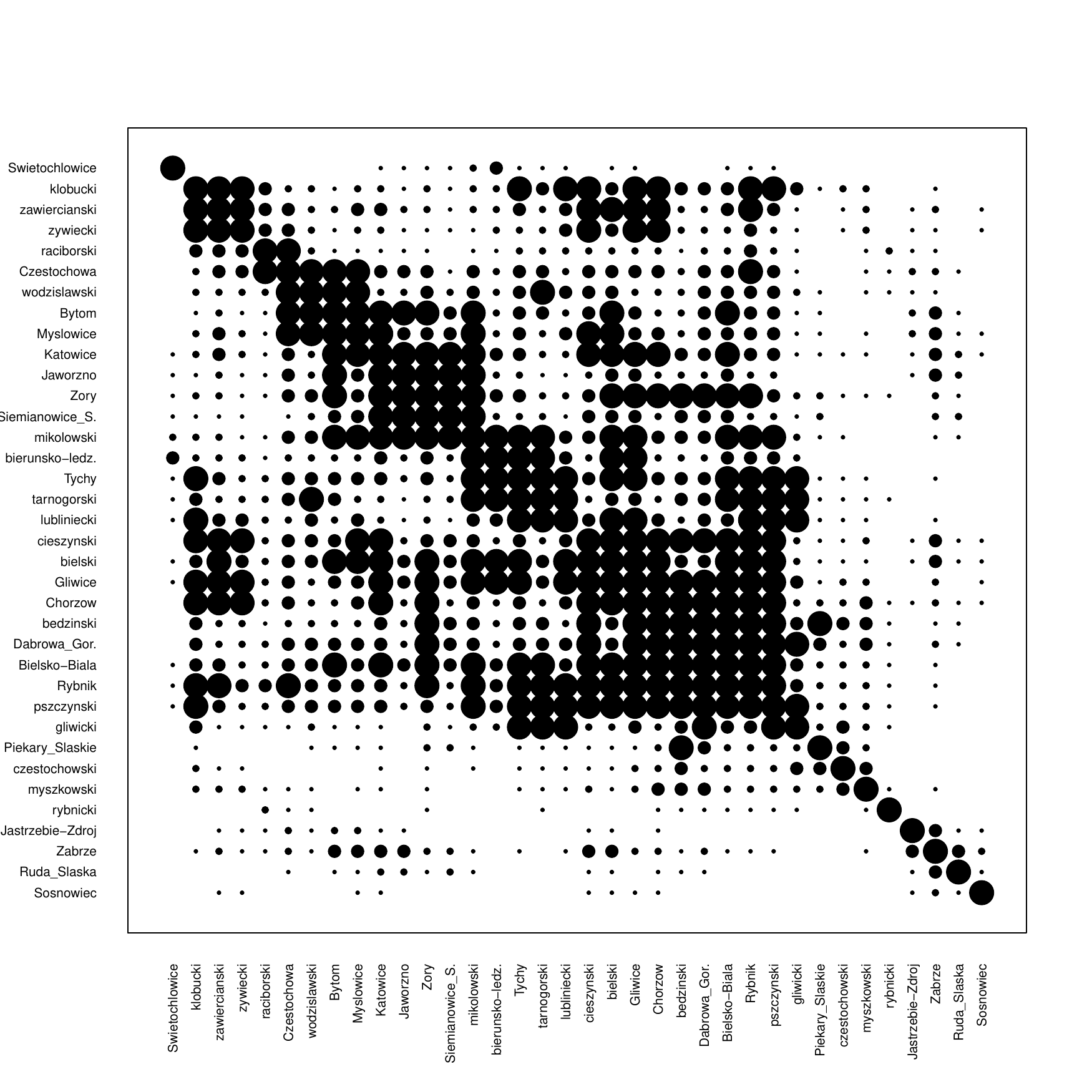} 
\includegraphics[width=0.32\textwidth]{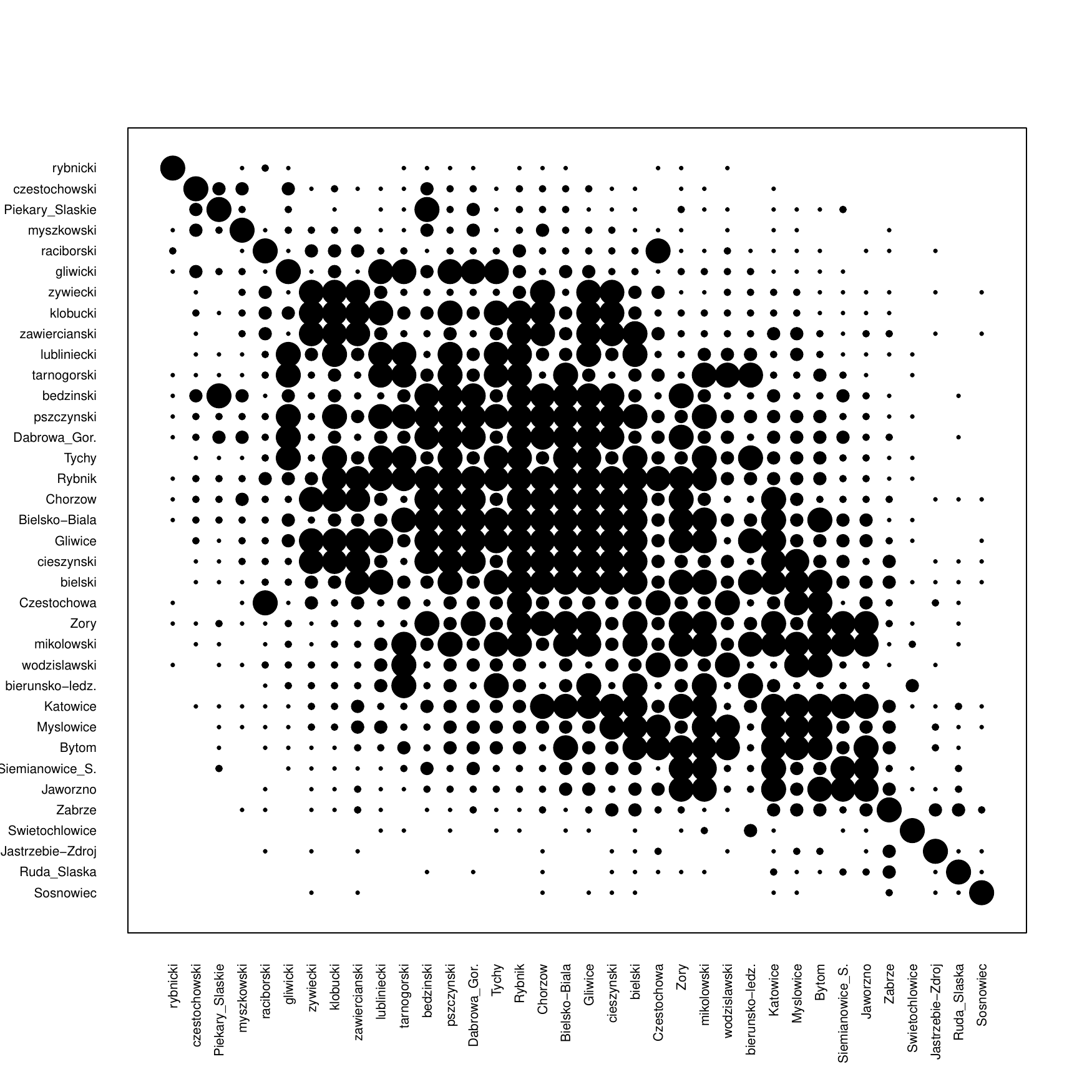}
\end{center}
\caption{
Visualization of \citepos{KWar2015} internet availability data. 
Top row: diagrams with \citepos{JCze1909}
original proposal presenting in each column the most similar rows,
bottom row: symmetric diagram directly visualizing the distance matrix.
Left column: data presented by \citet{KWar2015} found by \pkg{MaCzek} ordering
($U_{m} = 54.224$, path length = $64.021$, $U_{m}$ was minimized), center column:
diagram produced by \texttt{order="OLO"}
($U_{m} = 51.074$, path length = $59.63$, path length was minimized,
and right column:
diagram produced by \texttt{order="QAP\_2SUM"}
($U_{m} =46.769$, path length = $81.037$, $U_{m}$ was minimized).
\label{figKWar2015} 
}
\end{figure}

The ordering on the labels in Fig. \ref{figKWar2015} is of course difficult to read--off
the axes labels so we can display it directly
\\ \noindent
\code{>print(czkm\_internetavailability\_OLO)}\\ \noindent
\code{>print(czkm\_internetavailability\_qap2sum)}\\ \noindent
where the passed to \code{print()} objects 
\code{czkm\_internetavailability\_OLO} and \code{czkm\_internetavailability\_qap2sum} 
are the output of the ordering 
based respectively on the \code{"OLO"} and \code{"QAP\_2SUM"} methods. 
Hence, \citepos{KWar2015} original, \pkg{MaCzek} based, ordering of the counties is (with clusters identified by her)
\begin{itemize}
\item cluster I:  bielski, Gliwice, Rybnik, pszczy{\'n}ski, Bielsko--Bia{\l }a, D{\k a}browa G{\'o}rnicza, 
b{\k e}dzi{\'n}ski, Chorz{\'o}w, cieszy{\'n}ski,
\item cluster II:  Mys{\l }owice, Bytom, wodzis{\l }awski, Cz{\k e}stochowa, 
\item raciborski (singleton), 
\item cluster III: {\.z}ywiecki, zawiercia{\'n}ski, k{\l }obucki,
\item cluster IV: lubliniecki, Tychy, tarnog{\'o}rski, 
\item cluster V: miko{\l }owski, Katowice, Jaworzno, {\.Z}ory, Siemianowice {\'S}l{\k a}skie, 
\item unclustered singletons: bieru{\'n}sko--l{\k e}dzi{\'n}ski, 
{\'S}wi{\k e}toch{\l }owice, gliwicki, cz{\k e}stochowski, myszkowski, Piekary {\'S}l{\k a}skie, 
rybnicki, Zabrze, Jastrz{\k e}bie--Zdr{\'o}j, Ruda {\'S}l{\k a}ska, Sosnowiec,
\end{itemize}
the ordering under the \code{"OLO"} method is 
\begin{itemize}
\item {\'S}wi{\k e}toch{\l }owice (singleton), 
\item cluster III: k{\l }obucki, zawiercia{\'n}ski, {\.z}ywiecki, 
\item raciborski (singleton), 
\item cluster II: Cz{\k e}stochowa, wodzis{\l }awski, Bytom, Mys{\l }owice, 
\item cluster V: Katowice, Jaworzno, {\.Z}ory, Siemianowice {\'S}l{\k a}skie, miko{\l }owski, 
\item bieru{\'n}sko--l{\k e}dzi{\'n}ski (singleton), 
\item cluster IV: Tychy, tarnog{\'o}rski, lubliniecki, 
\item cluster I: cieszy{\'n}ski, 
bielski, Gliwice, Chorz{\'o}w, b{\k e}dzi{\'n}ski, D{\k a}browa G{\'o}rnicza, Bielsko--Bia{\l }a, 
Rybnik, pszczy{\'n}ski, 
\item unclustered singletons: gliwicki, Piekary {\'S}l{\k a}skie, cz{\k e}stochowski, myszkowski, rybnicki, 
Jastrz{\k e}bie--Zdr{\'o}j, Zabrze, Ruda {\'S}l{\k a}ska, Sosnowiec
\end{itemize}
and under the \code{"QAP\_2SUM"} method
\begin{itemize}
\item unclustered singletons: rybnicki, cz{\k e}stochowski, Piekary {\'S}l{\k a}skie, myszkowski, raciborski, gliwicki, 
\item cluster III: {\.z}ywiecki, k{\l }obucki, zawiercia{\'n}ski, 
\item cluster IV (part): lubliniecki, tarnog{\'o}rski, 
\item cluster I (part): b{\k e}dzi{\'n}ski, pszczy{\'n}ski, D{\k a}browa G{\'o}rnicza, 
\item cluster IV (part): Tychy, 
\item cluster I (part): Rybnik, Chorz{\'o}w, Bielsko--Bia{\l }a, Gliwice, cieszy{\'n}ski, bielski, 
\item cluster II (part): Cz{\k e}stochowa, 
\item cluster V (part): {\.Z}ory, miko{\l }owski, 
\item cluster II (part): wodzis{\l }awski, 
\item bieru{\'n}sko--l{\k e}dzi{\'n}ski (singleton), 
\item cluster V (part): Katowice, 
\item cluster II (part): Mys{\l }owice, Bytom, 
\item cluster V (part): Siemianowice {\'S}l{\k a}skie, Jaworzno, 
\item unclustered singletons: Zabrze, {\'S}wi{\k e}toch{\l }owice, Jastrz{\k e}bie--Zdr{\'o}j, Ruda {\'S}l{\k a}ska, Sosnowiec.
\end{itemize}

We can see that the minimization of the Hamiltonian path length resulted in the exact same clusters
that \pkg{MaCzek} identified. The singleton counties were differently placed and ordered (however, e.g.
the raciborski county is between clusters II and III). The ordering between the clusters and inside the clusters
is also different but these differences would seem rather minor. Actually if one looks at the diagram in
Fig. \ref{figKWar2015} it would seem that the \code{"OLO"} method tries to additionally build a cluster from
clusters II and V. Such a cluster could have some support as it would be predominantly made up of
large cities with county status. On the other hand, the \code{"QAP\_2SUM"} method, even though it has
a lower $U_{m}$ factor than both \pkg{MaCzek}'s output and the \code{"OLO"} method, breaks up all the clusters
bar cluster III. The graph seems to suggest that it attempts to construct one big cluster and drawing
any conclusions from it seems difficult. In this case the asymmetric diagrams do not seem useful. Even
if one varied the values of the \code{column\_order\_stat\_grouping} parameter, 
increasing the amount of plotted points to e.g. \code{column\_order\_stat\_grouping<-c(8,10,12,16)}, 
interpretable clusters did not start to appear. From the definition of the 
asymmetric method it seems that it has difficulties in usefully presenting data that has multiple
clusters of varying sizes (unlike in e.g. Fig. \ref{figASol2000} where all the clusters were
of similar size).

\end{section}
\begin{section}{Discussion}\label{secDisc}
In all of the run analyses it is important to point out that the user might obtain
a different ordering when rerunning the code. This was especially noticed when minimizing
$U_{m}$ through the \code{"QAP\_2SUM"} method. Two runs (for original and symmetric diagrams) 
often resulted in slightly different orderings, the one with better $U_{m}$ was chosen. All of the studies
were repeated a number of times and the re--run with the best values of the criteria
was retained (with a save of the random seed, provided in the Supplementary Material).
It can also happen that more than one ordering can have the same path length, especially
when there are more observations. In such cases perhaps another criterion, e.g. $U_{m}$, should be used
to differentiate between the best orderings. This took place in the analysis of \citepos{ASol2000} 
seals data. There were two arrangements that differed in the relative ordering
of observations (but these were jointly in the same place in both
orderings)---Fig. \ref{figASol2000}. 
However, the ordering visualized by the symmetric diagram had a slightly better value of $U_{m}$, than the 
one presented in the asymmetric diagram.

In general \citepos{JCze1909} original asymmetric diagram seems to provide a 
better visualization of the similarity of the ordered observations
in the example analyses where all the clusters are of similar size. 
However, this was as there were only six non--blank symbols 
per column. Hence, a significant amount of information is lost---and we are not distracted
by spurious similarities, if the main ones are focused around the diagonal.
On the other hand when considering \citepos{ASolPJas1999} urns data
or \citepos{ASol2000} seals data it would seem that the clusters visualized 
by the asymmetric diagram (with the default setting for the symbols) are
too small, and the symmetric diagram better captures their sizes.
Such an issue is even more evident with \citepos{KWar2015} internet availability
data. The asymmetric diagram fails to capture the grouping in an evident manner,
this happens even when the \code{column\_order\_stat\_grouping} parameter
was manipulated (so that it had a chance of corresponding to what the symmetric diagram was presenting). 
However, the reason for this could be that there were many clusters of varying sizes,
while the asymmetric method takes the same number of most similar elements for each column
(ignoring the magnitude of the similarities).

Furthermore, the \code{"OLO"} method, minimizing Hamiltonian path length,
seemed to outdo $U_{m}$ minimization. This was particularly evident in the simulation
study, in Section \ref{secObj}, when the seriation method, managed to group
objects into their correct clusters more successfully than the other two,
especially with difficult setups. The re--analysis of \citepos{KWar2015} internet availability
data makes this conclusion even more evident. Using \pkg{MaCzek}, whose genetic algorithm
minimizes $U_{m}$, \citet{KWar2015} found an ordering that graphically divided the counties
into five clusters (and left a number of singletons). The \code{"QAP\_2SUM"} method, 
found (as it usually does)  a lower value of $U_{m}$, but broke up the five clusters 
and instead attempted to create one big cluster that is ``huddled'' as much as
possible around the diagonal. Such a behaviour can also be seen in Fig. \ref{figphylQAP2SUMOLOGA}.
As already mentioned, and as \citet{AVas2019} observed, minimizing $U_{m}$ 
tends to work by forcing objects that have a large distance
between each other to be far away. But when ordering data, what we actually want is 
that close--by objects should be clustered together and Hamiltonian path--length minimization
seems to be doing this. Interestingly, even though \pkg{MaCzek} minimized $U_{m}$, its 
genetic algorithm seems to settle at a local minimum, whose ordering is similar to the one
minimizing the path length.

\end{section}

\section*{Acknowledgements}
We would like to thank two anonymous Reviewers for their comments that greatly improved
this work. KB is supported by the Swedish Research Council (Vetenskapsr\aa det) grant no. $2017$--$04951$.
We are very grateful to Ewa {\L }{\k a}czy{\'n}ska--Bartoszek
for translating \citepos{JCze1909} article from German, allowing 
us to implement his original method, as described in Section \ref{secCzkDiag}
and redo his skull study (Section \ref{secAnalyses}).
We would like to thank Arkadiusz So{\l }tysiak for many helpful comments and providing the seals and urns data,
Katarzyna Warzecha for providing the Internet availability data
and Miros{\l }aw Krzy{\'s}ko for his initial encouragement for creating an \proglang{R}
version of \pkg{MaCzek} at the XXIV National Conference 
Applications of Mathematics in Biology and Medicine in Zakopane--Ko{\'s}cielisko, $2018$.
The original version of \pkg{RMaCzek} was implemented as part of AV's 
master thesis in Statistics and Machine Learning
``Czekanowski's Diagram: Implementing and exploring Czekanowski's Diagram with different seriation methods''
($2019$)
done at the Division for Statistics and Machine Learning, 
Department of Computer and Information Science, Link\"oping University.

\noindent
\textbf{Supplementary Material}
\\ \noindent
The random seeds and source code required to obtain the results in the manuscript can be found in the 
repository \\ \url{https://github.com/krzbar/RMaCzek_BiometricalLetters_2020} .

\bibliographystyle{plainnat}
\bibliography{RMaCzek}

\end{document}